\documentclass{article}

\usepackage{graphicx}%
\usepackage{multirow}%
\usepackage[table]{xcolor}%
\usepackage{booktabs}%

\usepackage[T1]{fontenc}

\usepackage{subcaption}
\usepackage{threeparttable}
\usepackage{longtable}

\usepackage{hyperref}
\usepackage{todonotes}

\usepackage{enumitem}

\usepackage{longtable}
\usepackage{natbib}
\usepackage{comment}

\usepackage{multibib}
\newcites{review}{~}

\usepackage{authblk}

\begin{document}

\title{What do we study when studying politics and democracy? A semantic analysis of how politics and democracy are treated in SIGCHI conference papers}

\author[1,2,*]{Matti Nelimarkka}
\author[2]{Ville Vuorenmaa}
\affil[1]{University of Helsinki, Finland}
\affil[2]{Aalto University, Finland}
\affil[*]{Corresponding author: matti.nelimarkka@helsinki.fi}

\maketitle

\begin{abstract}
Human--computer interaction scholars are increasingly touching on topics related to politics or democracy.
As these concepts are ambiguous, an examination of concepts' invoked meanings aids in the self-reflection of our research efforts.
We conduct a thematic analysis of all papers with the word `politics' in abstract, title or keywords ($n$=378) and likewise 152 papers with the word `democracy.'
We observe that these words are increasingly being used in human-computer interaction, both in absolute and relative terms.
At the same time, we show that researchers invoke these words with diverse levels of analysis in mind:
the early research focused on mezzo-level (i.e., small groups),
but more recently the work has begun to include macro-level analysis (i.e., society and politics as played in the public sphere).
After the increasing focus on the macro-level, we see a transition towards more normative and activist research, in some areas it replaces observational and empirical research.
These differences indicate semantic differences, which -- in the worst case -- may limit scientific progress.
We bring these differences visible to help in further exchanges of ideas and human--computer interaction community to explore how it orients itself to politics and democracy.
\end{abstract}

\section{Introduction}

Human--computer interaction researchers have been active in studying and discussing the implications of technology for politics and democracy.
Various human--computer interaction research communities have been familiarised with this discussion through keynotes in key conferences, such as ``Could Social Media be a Threat to Democracy?'' (Ghonim, CHI 2017) and ``A discussion on media, platforms, and bias'' (Bell \& Angwin, CSCW 2018).
Beyond keynotes, various panels, special-interest groups, and workshops and active research shows human--computer interaction researchers engage with these topics and conduct scholarly work in them.
However, the \emph{meanings} of such words as `democracy' and `politics' are not self-evident.
These two concepts are essential to political scientists, but even they find the meanings of these words to be ambiguous \citep[more extensively, see Section~\ref{sec:definitions}]{Sigelman2006}.
Such blurriness may even lead us to speak past one another \citep{Abend2008}.
For example, for some `politics' is used the speak primarily about observational research focused on political actors, like elected representatives \citep[e.g.]{McDonald2019a}, while others may focus on research seeking to change the world in along the lines of post-colonialist and feminist research \citep[e.g.,]{Keyes2019}.
While both speak of politics, they seem to refer to different kind of research paradigm and activities -- and might disagree if the research focus on truly political phenomenon.

Therefore, conceptual ambiguity can limit our ability to properly communicate our knowledge, solve problems, and ultimately advance theory \citep{Oulasvirta2016,Abend2008}.
To explore this conceptual ambiquity, we must as \emph{semantic questions} on  the uses and meanings of individual words \citep{Abend2008}.
Within sociology, \citet{Abend2008} examines the meaning of the word `theory' by distilling seven different meanings of `theory' in sociological literature, ranging from a logically-connected system of propositions that establishes a relationship between variables to the studying and interpretation of prior authors, such as Habermas or Durkheim.
In human--computer interaction, \citet{Hornbaek2019a} similarly examined the meanings of the word `interaction' by analysing its modifiers in the CHI papers from 1981 to 2016.
They examined noun phrases that specified the interaction studied -- such as statistical interaction (e.g. `three-way interaction') or styles of interaction (`touch interaction', `cross-device interaction').
They identified a total of six different themes of modifiers, and further illustrated the range of modifier use regarding how we work with interactive systems.
These analytical approaches moved away from higher-level systematic literature reviews and focused on individual words and their meanings.
Therefore, this approach is fruitful in examining conceptual ambiguities and bringing conflicting meanings into academic discussion.

The driving challenge for this work is understanding of the uses of the words `\texttt{politics}' and `\texttt{democracy}' within human-computer interaction literature.
To explore this topic, we ask
\begin{itemize}
	\item \textbf{RQ1} what themes are addressed in conjunction with the words \texttt{democracy} and \texttt{politics} at SIGCHI conferences
	\item \textbf{RQ2} what temporal patterns reveal about the changing meanings of these terms
\end{itemize}
The paper first briefly reviews a political science perspective on these concepts
and then describes the data and methods used to conduct the analysis, from the selection of the material to analytical approaches,
describes the data and answers our research questions.

Following our empirical analysis, we unpack our results by identifying different orientations towards politics and democracy.
We identify a difference in their orientation towards politics and democracy; for some scholars, their study is the object of the research, while for others, it is an instrument to achieve their research goal, which is outside politics or democracy.
Similarly, we observed that orientations differ in their level: scholars examine a phenomenon through a mezzo-, or macro-level lens, or perceive it as out-of-reach.
Finally, our analysis highlights the differences in researcher positionality: a phenomenon can be either observed or influenced.
These different orientations speak of the importance of a plularistic meaning for democracy and politics.
However, our analysis suggests that research efforts are more focused on certain orientations.
To ensure plularity, we invite scholars to ensure the breath of the research agenda,
and consider review and evaluation practices.
We also ask how human--computer interaction researchers learn to attach politics and democracy to certain research topics:
this may produce blind spots in our work, and different attachments seem to drive the themes and research approaches.
As the importance of politics and democracy is increasingly being acknowledged in the technology sector, we are inviting us to carefully think about how our scholarship speaks to society.
Our work provides an overview of the field, suggests some common ground, and ensures the relevance and rigor of our ongoing research.

\section{What politics and democracy are: A political science perspective}
\label{definition:politics_democracy}
\label{sec:definitions}


Before examining how human--computer interaction researchers have approached democracy and politics,
we examine how the term is approached in political science, the discipline, whose core focuses include politics and democracy.
The question is anything but simple: even within a single field (e.g. political science), these terms are blurry \citep[for example][]{Sigelman2006}.
For example, those researching political activism, a subtheme within political science, have found it difficult to specify what \emph{political} activities are.
In his paper ``Studying Political Participation: Towards a Theory of Everything,'' \citet{van2001studying} shows that a sample of just six papers on political participation identified nearly 70 distinct activities (``voting,''
``deliberately bought certain products,''
``abstained from voting out of protest,'' etc.).
This sample displayed a shift in focus over time, from elections to campaigning, contacting political officers, political protests, and lately even civic engagement in general.
His main argument is about the challenge of understanding political participation (and, through this, what politics is): it has various meanings even within studies centered on political activism.
His observations on the challenges of separating politics from non-politics have been echoed by others in the field \citep[e.g.,][]{McAuley2018,Shively2011}.

Several introductory textbooks\footnote{~According to \citet{platt1998history}, textbooks articulate the consensus of a discipline and demonstrate its shared positions.
Furthermore, she argues that because textbooks are so widely used, researchers are familiar with them and represent their content, even implicitly, in their research practises.
Our choice to use textbooks to construct our argument reflects this important position in actually defining disciplines.} help students embarking on political science to navigate the complex terminology and offer working definitions of politics.
In these books, the definitions of politics range from
suggesting that it is ``the activity by which groups reach binding collective decisions through attempts to reconcile differences among their members'' \citep{McAuley2018}
or that it has to do with ``making a common policy for a group'' and ``exercising power by one person or persons over another person or persons'' \citep{Shively2011} to
efforts to extend the definition beyond public governance.
Textbooks following the latter approach state
that politics is about developing and maintaining normative values and identities in a nation-state, and hence the creation of boundaries \citep{Carr2007},
or that politics ``encloses and involves many areas of social life, such as social class, ethnicity, gender, [and] identity'' \citep{McAuley2018}.
It is clear that even the textbooks do not posit a ``one-size-fits-all'' definition.

An established stream of critical scholarship in political science calls for expanding the consideration of ``politics'' beyond society-level elements such as exercising power in society or making collective decisions as a society (the politics of institutions and society, often referred to as ``politics with a capital `P'").
For example, from a feminist perspective on politics, the personal is also political.
The aim of such work has been to widen the boundaries of what is deemed politics (and political science), to address private and social matters as well \citep[e.g.,][]{bryson2003feminist}.
With the latter delineation of politics (covering what has been termed ``politics with a small `p'"), the political is implicit and explicit in such domains as family life; politics is at the core of these, just as much as it is at the core of public decision-making.

More importantly, political science is a pluralistic field, not driven toward any single theory or framework \citep{Almond1988}.
Rather, as articulated by \citet{Sigelman2006}, ``political science has always been a federation of loosely connected subfields rather than a tightly integrated field of study.''
For this reason, the political science community actively seeks to reflect on the state of the discipline and consider how it should be developed \citep[e.g.,][]{easton2002development,TheStateoftheDisciplinetheDisciplineoftheState}.
In doing so, it highlights the benefits of a pluralist approach and dialogue spanning various positions \citep[e.g.,][]{marsh2002theories}.
Recognising the persistent plurality of thought in political science, we could easily extend the reflection on the various definitions, both competing and complementary, but that would be beyond the scope of this section.
The brief review above indicates well enough that a definition of politics encompasses social and cultural contexts \citep{McAuley2018}.

The ``democracy'' aspect of our thematic analysis presented challenges similar to those identified for ``politics.''
Discussing the differences between representative and direct democracy, \citet{Becker2001} proposes that technology ought to transform democracy from a representative function into a direct, citizen-led enterprise.
Among the others examining the various forms that democracy can take is
\citet{Dahlberg2011}, who suggests that online democracy can be understood in four ways:
a) as registration and subsequent aggregation of choices (in the liberal-individual model);
b) as existing in rational and respectful discussions in which opinions are formed (in the deliberative model);
c) as supporting the formation of political groups and activism (in the counter-public model);
or
d) as helping citizens bypass the anti-democratic centralised state and capitalist systems (in the autonomist Marxist model).
Even this relatively short list highlights variety, in that the concept need not be limited to the form of decision-making (as it is in the first two models); it can also include ideas about governance and economic contexts (as in the counter-public and autonomist Marxist models).
A classic of political science, \emph{Models of Democracy} by \citet{held06}, identifies a dozen separate democratic models, or configurations of decision-making rules, economic structures, and understandings of the ``stakeholders'' or citizens in the democratic system.
These examples alone demonstrate that ``democracy'' too is a complicated term, which is true also in human--computer interaction \citep{Nelimarkka2019a}.

Even this \emph{brief} review has shown that ``politics'' and ``democracy'' are complicated concepts:
we limited it to discussion by political scientists, and still no commonly agreed nomenclature emerged.
As is alluded to above, the definition hinges partially on authors' position and socio-cultural background \citep{McAuley2018}.
For example, the realm of politics may be extended to personal spaces in the hope of drawing attention to decision-making in private, as is done by feminist political science theory \citep{bryson2003feminist}.
Therefore, it is not surprising if human--computer interaction researchers use these phrases in a diverse manner.
However, illustrating the diversity of semantic use may help us to acknolwedge biases in our research efforts and shape the future research agenda.


\section{Data and methods}

To answer our semantic question -- the meanings attached to politics and democracy -- we select a corpus representing `politics' and `democracy' \emph(at face-value): using the ACM Digital Library\footnote{ We acknowledge that the sample does not include, for instance, the \emph{European Conference on Computer-Supported Cooperative Work} or \emph{International Journal of Human--Computer Studies} articles.
However, SIGCHI conferences, such as CHI and CSCW, are at the forefront of publishing on human--computer interaction and should be representative of the state of the discipline in relation to the semantic question.
Obviously, limitations arise when the focus is restricted to ACM SIGCHI conferences.
For example, ACM SIGCHI conferences have a North-American bias \citep{dePaula:2015:DCR:2685553.2685560}.
This must be borne in mind when the results are interpreted: the papers represent not human--computer interaction work in general but a subset of what is published in connection with this set of conferences.
We acknowledge and discuss this aspect further in limitations.
}, we identified all texts that used these words on their titles, keywords or abstracts.
This demonstrates how authors currently use these words to develop their arguments: if authors used these words, we included their work in the review.
To illustrate, \citet{10.1145s1385569.1385652} studied a democratic jukebox for selection of music at parties.
We included their paper in our analysis, even though there may be \textit{limited} wider societal implications in the partygoers' music selection, and hence the work may be far removed from the traditional sphere of political science.

Collected material was then analysed in detail to understand themes in relation to our keywords.
Our research questions are focused on understanding the semantic word use, therefore we employ a thematic analysis of the literature \citep[e.g.,][]{Bossen2017,Suominen2016,Griffiths2004}.
However, we do not focus on the themes of the research, but on the meanings of \texttt{politics} and \texttt{democracy}.
Therefore, our scope is different from traditional systematic literature reviews, which offer a synthesis of the literature \citep[e.g.,][]{Gross2013}, while others classify and cluster scholarly works on the basis of pre-existing coding schemata for such factors as the methods used \citep[e.g.,][]{Dillahunt2017,Wallace2017}.

The analysis was based on an inductive analysis by two coders and was further examined by triangulating the results with unsupervised machine learning.
In the following sections, we describe the process in more detail.


\paragraph{On the scoping of politics and democracy}


\begin{figure}

    \centering
    \includegraphics[width=0.75\textwidth]{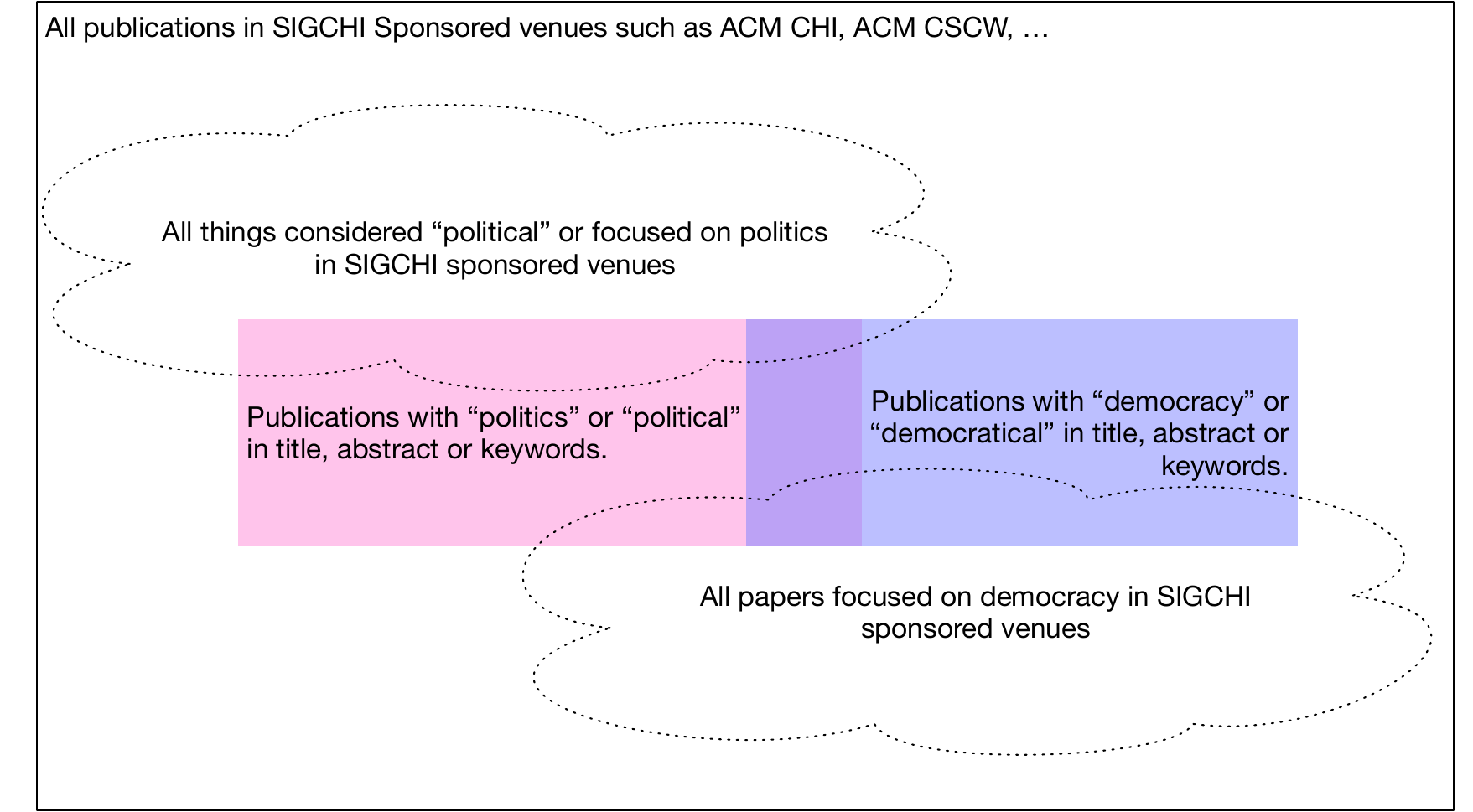}
    \caption{Semantic literature review and its relation to substantive domains.}
    \label{fig:semanticliteraturesearch}
    
\end{figure}

Focus on exact occurrences for `politics' and `democracy' leave material which somehow touches \textit{political} or \textit{democratic} topics outside (see Figure~\ref{fig:semanticliteraturesearch}).
Those who focus on institutional politics, or capital P-politics, might suggests adding words such as voting and elections to more comprehensively capture these.
Others might highlight the important role of social movements and civic society and suggest those as additional keywords.
Similarly, scholars currently doing critical work in human--computer interaction, such as works around social justice \citep[for a recent review, see][]{Boone2024}, might suggest addition of versatile terms to the lists, such as feminism, post-colonialism, etc.
Others might indeed do justifiable arguments that democratic choice of music is not really about democracy, even while it uses this term.

Therefore, it is unclear how to define the boundary on what \textit{is} and what \textit{is not} political (or democratic) -- as Section~\ref{sec:definitions} suggests, even the field of political science does not have a clear answer for this.
In the broadest definition of politics, any evaluation criteria means there must ne values and therefore, it is somehow political \citep{Weiss1993}.
From this perspective, even keyboard layout optimisation -- focused on speed and accuracy, not for example depth of thinking -- could be framed as a question of politics.
Therefore, the scoping of the literature is a slippery slope, and given the contested nature of politics, it would is a political decision to choose what conceptually to include and exclude.
To address this challenge, the analysis focuses only on the semantic meanings and face value of these terms.

\subsection{Execution of the literature selection}

\begin{figure}
	\centering
	\includegraphics[width=\columnwidth]{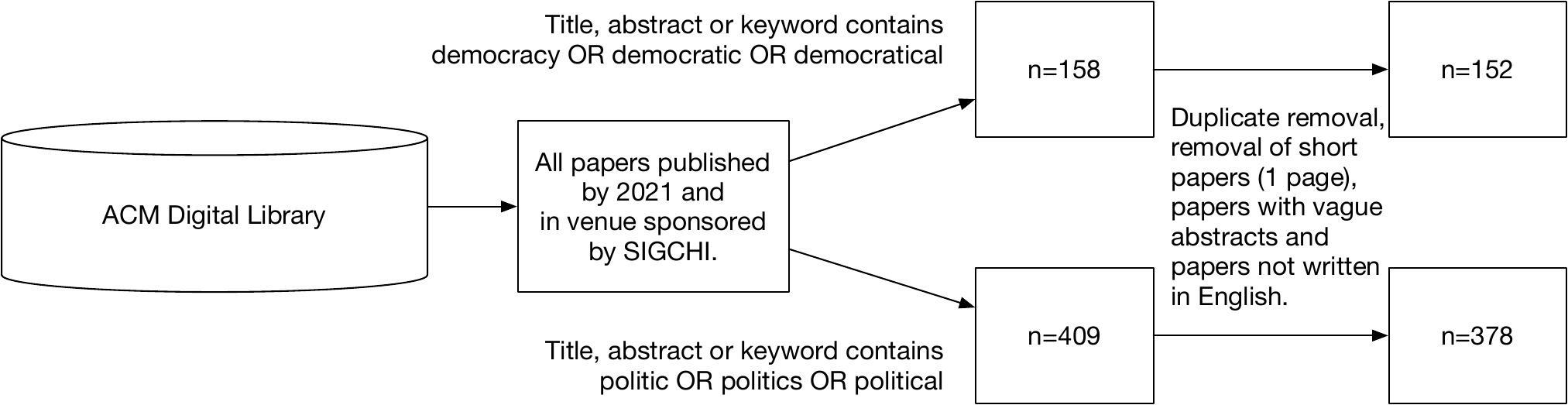}
	\caption{Literature selection was based on searching of keywords and conducting a minimal exclusion by removing duplicates and material which could not be classified.}
	\label{fig:selection_flow}
\end{figure}

\paragraph{Inclusion criteria:}
We used the ACM Digital Library interface to search for the terms politic, politics, political and democracy, democratic, democratical.\footnote{ Using queries \texttt{"query": {Abstract:(politic OR politics OR political) OR Keyword:(politic OR politics OR political) OR Title:(politic OR politics OR political)}
"filter": {Sponsor: sigchi},{Publication Date: (01/01/1908 TO 12/31/2020)},{NOT VirtualContent: true}
} and \texttt{"query": {Abstract:(democracy OR democratic OR democratical) OR Keyword:(democracy OR democratic OR democratical) OR Title:(democracy OR democratic OR democratical)}
"filter": {Sponsor: sigchi},{Publication Date: (01/01/1908 TO 12/31/2020)},{NOT VirtualContent: true}
}.
Search with different words sought to ensure the words are captured even after stemming, which ACM Digital Library uses in its textual search.
}
The results were limited to publications presented at conferences sponsored by SIGCHI, such as CHI, CSCW, and DIS, and published before 2021, and only to material where these keywords appear in the title, abstract or keywords.
The literature collection was conducted on January 28th, 2021.
In total, this yielded 409 articles on politics and 158 articles papers on democracy for potential inclusion in the analysis (see Figure~\ref{fig:selection_flow}).

\paragraph{Exclusion criteria:}

We excluded duplicates and papers that did not specify the details necessary for our classification procedure: for example, some works consisted of one page of content, had such a vague abstract that it was unclear what the authors had done, or were not written in English.
We intentionally kept the material inclusive not to bias our findings based on some pre-existing notation of politics or democracy: the excluded material mostly contained of duplicates.
The final sample consisted of 378 papers for politics and 152 for democracy.

\subsection{Analysis of the literature}

Our work focused on the semantic meanings for words politics and democracy.
This work is not a classical literature review seeking to map out all work under which relates to politics or democracy, but only on the literate use of these terms (see Figure~\ref{fig:semanticliteraturesearch}).
Methodologically, there are different ways of working with such questions:

For example, \citet{Abend2008} identifies seven different meanings for the word theory in sociology.
Among these seven different ways are including theory to account for fundamental normative components (such as post-colonial theory or feminist theory) or re-interpreting classical texts in sociology (such as Marx, Weber, Durkheim, Simmel, Parsons, Habermas, or Bourdieu) or that theories explain causal claims.
As these three examples illustrate, the use of the term is diverse among sociology community.
However, while insightful and opens conversations, his account is not empirical -- it is based on reflection and experience in the field.

More empirical account can study the occurrences of individual words. For example, in human-computer interaction, \citet{Hornbaek2019a} studied the meanings of the word interaction through its modifies.
Their work highlights changing technology landscape as interaction was modified with words such as direct, physical, multitouch, mouse, screen.
With modification, they refer to examining the noun phrases related to interaction from full-paper publications from CHI in 35 years.
For example, in the sentence ``With increased interest in touch screen interaction, \ldots'' the word interaction is modified with the words touch screen.
They further group these identified words based on how they conceptualise interaction.
While this is a clear empirical approach, the meaning of politics and democracy -- allowing us to deduce the context why it is used -- may not be apparent from modifiers.
For example, in the example below , the focus of authors is not on conservative political climate, but rather it focuses on bringing scholars together to discuss orgnanisation of the conference in a context which is perceived to also have conservative political climate:

\begin{quote}
    ``In this alt.chi paper, we reflect on \#CHIversity a grassroots campaign highlighting feminist issues related to diversity and inclusion at CHI2017, and in HCI more widely. \#CHIversity was operationalised through a number of activities including: collaborative cross-stitch and 'zine' making events; the development of a 'Feminist CHI Programme'; and the use of a Twitter hashtag \#CHIversity. These events granted insight into how diversity discourses are approached within the CHI community. From these recognitions we provide examples of how diversity and inclusion can be promoted at future SIGCHI events. These include fostering connections between attendees, discussing 'polarizing' research in a conservative \underline{political} climate, and encouraging contributions to the growing body of HCI literature addressing feminisms and related subjects. Finally, we suggest how these approaches and benefits can translate to HCI events extending beyond CHI, where exclusion may routinely go undetected.`` \citep{10.1145s3170427.3188396}
\end{quote}

\begin{figure}
	\centering
	\includegraphics[width=\columnwidth]{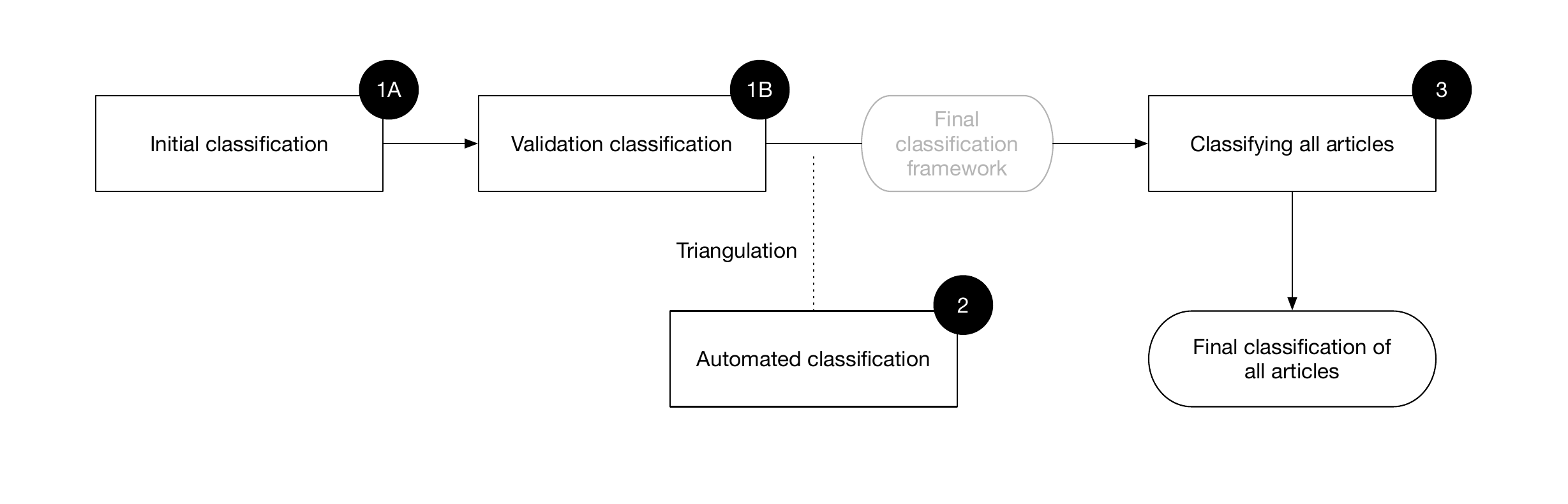}
	\caption{The four stages of the literature classification process -- development and refinement of the codebook (the first two stages), automated classification (the third stage), and creation of the final classification framework (the final stage).}
	\label{fig:classification_flow}
\end{figure}

This suggest that we cannot use a mechanic approach like this to explore these meanings.
While there are even more complex computational linguistic approaches to examine changing word context \citep[e.g.]{Hamilton2016}, these approaches start to require such extensive datasets that they are not suitable for them.
Therefore, we classified the literature via four separate stages, three manual and one based on automated text analysis (see Figure~\ref{fig:classification_flow}). 

In Stage 1A, we iteratively coded both corpuses separately.
During this stage, we developed the theme-based categorisation in accordance with the articles' titles and abstracts, focused on the context and use of the terms \texttt{politics} and \texttt{democracy}.\footnote{ To keep the task manageable, researchers commonly focus only on titles and abstracts in this type of thematic analysis \citep[e.g.,][]{Suominen2016,Griffiths2004}.
When in doubt, we opened the full text version of the text to assist in the analysis.}
We focused our attention to the semantic question: why the term was used in the titles and abstracts?
Due to this, we focused focused heavily on the word's surrounding sentences to understand the context of use further.
The two coders regularly met during the analysis process to discuss the work. Initially we examined emerging patterns and observations and inductively worked to develop potential categories while at the same time familiarising ourselves with the material \citep{Braun2006}.
At later stages, we focused more explicitly to discuss our coding decisions and observe and address differences in them to further clarify previously established categories.
It was during this stage that we determined that the categories should be mutually exclusive; i.e., each paper was assigned to only the most suitable category.
This decision is discussed in connection with Stage 2, in which we considered the possibility of placing a paper in multiple categories.

After this, in Stage 1B, we examined each category separately and ensured that the codes were applied coherently after initial categorisation.
We started the second round of coding with the classification schema developed in Stage 1A, which was further clarified on boundary cases.

Stages 1A and 1B were based on manual classification by a research assistant in collaboration with the lead author.
Both independently classified the material and discussed the disagreements.
Because of the complexity and interdisciplinary nature of the classification task,\footnote{ \label{footnote:validation}
In this type of classification, intercoder reliability analysis is often advisable. 
However, it is not common for literature reviews published in human--computer interaction, most likely because of the challenges of conducting said analysis properly in an interdisciplinary field like human--computer interaction.
Often, the topic is not addressed at all \citep[e.g.,][]{Dillahunt2017},
or the scope is rather limited.
For example, while \citet{Wallace2017} used two classifiers, they stated that ``[w]hen agreement could not be reached, [more senior] Rater 2's classifications were reported.''
This \emph{de facto} single-classifier approach illustrates that when two raters are used, a deep understanding of the domain is readily accorded greater importance.
While \citet{FrichPedersen2018} did calculate intercoder reliability, they claimed, after finding only moderate $\kappa$ values, that the ``ability to replicate each single categorization based on the final table is deemed reassuring.''
This suggests that even they did not find intercoder reliability analysis critical.} we deemed this approach warranted.

To address the important validity-related questions that arise with manual classification (stages 1A and 1B), we conducted methodological triangulation.
In Stage 2, topic models \citep[e.g.,][]{Blei2010} was used to classify the articles, i.e., develop a separate version of potential themes in the materials.
They to produce thematic mappings via analysis of abstracts and of full scientific papers \citep[e.g.,][]{Griffiths2004,Suominen2016}.
This methods extracts topics from document sets, i.e., statistically produced groups where some words are more frequently present in a subset of documents.
These are mixed-membership models; each abstract belongs to several topics.
This was useful as the method did not focus on our keywords but on the whole abstract, it is expectable that some of these topics emerge around particular methodologies and words related to them.
At the same time, it allows us to examine if our choice for the single-membership model in stages 1A and 1B.
We contend that if the classifications from these stages are similar to those from Stage 2, there is no significance in whether we assign a paper to a single category or several.
We used the Structural Topic Model software package, which was used to conduct standard textual preprocessing (such as removing stopwords and punctuation and stemming words) and conduct the topic modelling.
One key problem in topic modelling is choosing the number of topics the algorithm produces: STM has a build in low-dimensional embedding method \citep{mimno2014low}, which was used to choose the most fitting number of topics.
As \citet{Baumer2017a} have pointed out, topic models often produce a low-level classification of the data.
Therefore, we examined the output and, as necessary, merged topics into meaningful thematic clusters to set the triangulation's component classifications at the same conceptual level.
This can be thought of as an axial coding stage for the topic model work.

Finally, in Stage 3, all articles were coded once again (they had all already been coded twice, in Stage 1A or 1B and Stage 2), accounting the the final coding framework took into account the observations from stages 1A, 1B, and 2.

As detailed in footnote~\ref{footnote:validation}, validation in literature reviews, especially open-ended ones such as ours, is difficult.
Accordingly, in addition to the triangulation of methods, we validated the results by asking several scholars of social sciences and of human--computer interaction to provide comments on the manuscript, including the classification framework (see the acknowledgements for details).
Although it did not measure intercoder reliability, this step examined the classification's face validity.
Secondly, we have sought to make the sample selection and the classification process as transparent as possible.
For example, all papers are listen in Appendix~\ref{literature_reviewed} and classified on Appendix~\ref{classification_schema_2_details}, so readers may freely familiarise themselves with the literature.
This invites everyone to conduct further investigation and, for example, perform analysis with alternative search term selections or other classification procedures.

\section{Findings}

We have divided our findings into the following three sections:
the current section provides descriptive details on our literature,
Section~\ref{sec:rq1} provides a categorisation on the uses of politics and democracy as presented in our corpus,
and Section~\ref{sec:rq2} examines the temporal trends.
We intentionally kept these sections close to the data to allow readers to reflect on the findings without interpretative work.
We discuss this in more detail in the Discussion (Section~\ref{sec:discussion}).
We unpack how categorisations reveal different orientations scholars have towards politics and democracy.
We also demonstrate how these orientations change over time, and even observe how some orientations are omitted in contemporary research.
We also discuss how politics and democracy as concepts seem to generate different research questions for human--computer interaction researchers and propose how they could become more complementary.

As Table~\ref{tab:years} shows, the raw count of papers in our area of interest has clearly been increasing since the early 1980s.
In addition, the volume of these papers relative to all work for SIGCHI conferences has risen slightly and is at an all-time high.
Interestingly, our core topics were also prominent in the 1980s but experienced a dip in the 1990s to early 2000s, most likely contributed to increasing publication volume in human--computer interaction.
(Note that on early years, the absolute counts are low.
As the high relative counts indicate, the total number of publications was lower; in 2015--2020 there were about 88 times more publications than in 1980--1984.)
Then in the early 2010s there was a doubling of the publication output from the latter eras.
Naturally, at the same time the volume of all publications at SIGCHI conferences has increased.
For both corpuses we have computed the relative volume of publications compared to \emph{all} SIGCHI sponsored publications in the given timeframe, shown in column '\%'.
Overall, we show that the relative publication volume was relatively high in the early days of SIGCHI, showing that politics and democracy were used as words to describe scholarly output.
However, during the 1990s the relative volume decreased, indicating that scholars chose to use other terms to discuss this topic.
This downwards trend turned around in the 2000s, especially in the 2010s, and the words `politics' and `democracy' have increasingly been used to characterise the research.

\begin{table*}[t]

    \caption{Characterization of the publications.}

	\begin{subtable}{.45\columnwidth}

  \begin{small}

    \centering

    \begin{tabular}{lrrrr}
      \toprule
     & \multicolumn{2}{c}{\texttt{democracy}} & \multicolumn{2}{c}{\texttt{politics}} \\
     & \multicolumn{1}{c}{$n$} & \multicolumn{1}{c}{\%} & \multicolumn{1}{c}{$n$} & \multicolumn{1}{c}{\%} \\
      \midrule
      1980--1984 &   0 & 0.00 &  4 & 1.57 \\
      1985--1989 &   1 & 0.14 &  4 & 0.55 \\
      1990--1994 &   1 & 0.05 &  3 & 0.16 \\
      1995--1999 &   1 & 0.03 &  2 & 0.07 \\
      2000--2004 &   5 & 0.12 & 12 & 0.28 \\
      2005--2009 &  14 & 0.17 & 28 & 0.35 \\
      2010--2014 &  40 & 0.22 & 88 & 0.48 \\
      2015--2020 &  90 & 0.40 & 237 & 1.05 \\
      \midrule
      Total & 152 & & 378 & \\
       \bottomrule
    \end{tabular}


  \caption{A breakdown of the selected literature by five-year span.}
  \label{tab:years}

  \end{small}

\end{subtable}~\begin{subtable}[t]{.45\columnwidth}

  \begin{small}

    \centering
    \begin{tabular}{lrr}
      \toprule
     & \texttt{democracy} & \texttt{politics} \\
      \midrule
      CHI &  66 & 177 \\
      CSCW &  25 &  76 \\
      DIS &   6 &  26 \\
      HT &   1 &  14 \\
      UBICOMP &   7 &  11 \\
      TEI &   6 &   6 \\
      GROUP &   2 &   9 \\
      C\&C &   6 &   5 \\
      IUI &   7 &   4 \\
      CHI PLAY &   4 &   6 \\
       \midrule
      Other & 22 & 41 \\
       \bottomrule
    \end{tabular}

  \caption{The 10 most common venues for publication of the selected papers.}
  \label{tab:venues}

  \end{small}

\end{subtable}

\end{table*}

Accounting for about 60\% of the sample, CHI and CSCW were the most active publication venues for both corpuses.
These are shared by the various communities within the human--computer interaction field and represent high-quality output,
so this finding is not a surprise.
However, large contributions were also evident from outlets with a
focus on design (DIS) and
technology (IUI, UbiComp and HT) (see Table~\ref{tab:venues}).
We also see some differences between the keywords relative to corpus size:
for example, IUI and TEI conference papers use the word \texttt{democracy} more often while the HT conference clearly addresses \texttt{politics} more.
The high figures for top-tier venues such as CHI and CSCW are pleasing and highlight the importance of this kind of work for the community.

\section{RQ1: Thematic analysis of \texttt{democracy} and \texttt{politics}}
\label{sec:rq1}

Work on the first research question entailed characterising the literature on politics and democracy related to the SIGCHI conference.
Here, we will present the categories that emerged from our analysis and show the similarity of output between the manual and the automated classification.

\subsection{The manual classification}

The results of the manual classification illustrate the diversity and looming issues human--computer interaction researchers work when they use words politics and democracy.
We found a rich set of topics where scholars use these terms:
\begin{itemize}
\item the role of (computer-mediated) media in politics and democracy 
\item the importance of technologies to the functioning of civic society, ranging from participation to decision-making to the organisation and improvement of civic society.
\item politics and democracy in specific application contexts, such as workplaces and public services 
\item papers focused on methodologies either using political data or addressing the politics of methods 
\item papers broadly addressing the technological advancement in society, such as technology democratisation or broadly engaging questions regarding human-computer interaction from political or democratic stances 
\item papers focused on building an academic community, or papers which seek to motivate or contextualise their contribution from political or democratic perspectives or suggest implications for these categories 
\end{itemize}
In these wider groups, we can identify a total of 13 different themes (see Table~\ref{tab:rq1result}), illustrating different perspectives to study these six wider perspectives.
(For more examples and references to each paper, see Appendix~\ref{classification_schema_2_details}.)
Such diversity is in line with expectations, in light of the terms' various potential meanings as presented in Section~\ref{sec:definitions}.
As Table~\ref{tab:rq2counts} indicates, a double-digit number of papers was found for each class, suggesting that they are all large enough to merit being treated separately in the thematic classification.


\begin{longtable}{p{.1\textwidth}p{.65\textwidth}p{.22\textwidth}}

\\
    \caption{Classification schema: 13 different themes of politics and democracy}
    \label{tab:rq1result}
    \\
     \multicolumn{3} {l} {For more examples and references to each paper, see Appendix~\ref{classification_schema_2_details}.}
    \\ \toprule
Category & Description & Keywords \\
\midrule

\endfirsthead

\toprule
\multicolumn{3}{l}{\emph{Table \ref{tab:rq1result}: \nameref{tab:rq1result} continues.}} \\
Category & Description & Keywords \\
\midrule

\endhead

\midrule
\multicolumn{3}{l}{\emph{Continued on next page}} \\
\bottomrule
\endfoot

\bottomrule
\endlastfoot

\multicolumn{3}{l}{\textbf{Media}} \\
User generated content &
Research focused on user-generated content in media, either by examining current practices of political discussion or suggesting novel services to aid in producing and consuming content produced by other people.
The media is created ``by the many'' and is seen as dialogical in nature. Common data sources include Facebook, Twitter, Reddit, and comment sections on news media articles. & social media; Twitter, public sphere; Reddit; Facebook \\

Media consumption &
Studies on one-directional mass media consumption of television, radio, newspaper, and online modes of these, as well as consumption of user-generated content via blogs. The overall aim of these works is to improve media consumption, such as balancing news reading habits, identifying ``fake news'' or support in finding new perspectives on these topics. & news; credibility; selective exposure; Twitter; social media \\

\multicolumn{3}{l}{\textbf{Civic society}} \\
Public participation &
Research focusing on how technology supports public participation in society. Application areas range from public participation related to legislation proposals, city planning or allocation of (public) resources. &  civic engagement; deliberative democracy; urban planning; participatory democracy; participation \\
Social movements &
Studies on technology use by activists, social and political movements and social causes. Papers explicitly speak about social movements and emerging user needs within this context.  & social media; activism; Facebook; collectives; discourse \\
Social issues &
Human-computer interaction research exploring the intersection of social issues and technology. For example, papers have focused concretely on issues such as social integration of migrants through technology as well as higher-level academic discussion on issues such as the economy and how human-computer interaction researchers can address them.  & appropriation; social movements; solidarity economy; social media; computer club \\

\multicolumn{3}{l}{\textbf{Context of use}} \\
Public services &
Work discussing how technology could improve public services and infrastructure, e.g. by involving citizens in planning of services, improving access to services, providing opportunities for feedback or reducing costs.  & mobile participatory sensing; street sweeper; environmental justice; environmental science; air quality sensing \\

Organisation and workplace &
These papers focus on the use of technology in workplaces and other organisational settings. They use phrases such as ``office politics'' and ``workplace democracy'' to connect the theme of decision-making in an organisation to the words `politics' and `democracy'. These terms are catch-call words to express the social reality and power dynamics in organisations.  & cscw; informal interaction; medical records; organisational memory; shift change \\

\multicolumn{3}{l}{\textbf{Methodology}} \\
Design methods &
Work in the intersection of politics and design, including both the political nature of the design process and the democratic and political motivations that drive artifact design and design thinking. Unlike the above, the main driving motivation of the works is to advocate for improved ways of conducting design by better accounting for politics and democracy.  & design research, design; participatory design; social change; wellbeing; design fiction \\

Data analysis methods &
Papers developing new data analysis methods, either used on data that is political (such as political activity on social media) or the potential use of data analysis in a political context. Unlike in the works above, the main contribution is improved methodology.  & social media; sentiment analysis; social networks; user profiling; significance testing \\

\multicolumn{3}{l}{\textbf{Technology and society}} \\
Professional topics &
Exploration of the politics of human-computer interaction, both in reflective and critical stances towards computing research, such as lack of political engagement or regional representativeness of human-computer research. These papers focus on emerging issues in the human-computer interaction research community and reflect the political processes of concluding our research.  & activism; feminism; ethics; design; social change; hci education \\
Technology democratisation &
Discussion of democratisation of the technology, accessible design and infrastructure, and prototyping methods, motivated by increasing access to technology.  & identity; visualisation; diy; hci4d;scalable architecture \\

\multicolumn{3}{l}{\textbf{Other}} \\
Motivating context or background &
Paper uses democracy or politics to further motivate its aims, contextualise the study or highlight its potential implications. There is no clear connection to other specific policy topics presented above.  & ethnography; design; research though design; social media; transnational \\
Academic community activities & 
Various activities at the conference related to community building and networking, such as doctoral consortiums, workshops and special interest groups. Also including introductions written for proceedings organised by conference.  & social media; design; participation; participatory design; activism  \\

\end{longtable}

The diversity is also illustrated in the five most common keywords for each category, shown in Table~\ref{tab:rq1result}.
These keywords demonstrate the differences between the categories.
Even with the overlaps, the complementary keywords illustrate the difference.
For example `social media' is a keyword in the categories of user-generated content, social movements, social issues and data analysis methods.
However, other common keywords, such as `public sphere' in social media, 'collectives' in social movements, `solidarity economy' in social issues, and `sentiment analysis'  in data analysis methods indicate the substantial nuances between these categories.
Therefore, we find the categorisation to various themes merited in this case even while some of the keywords are reused.
Similarly, the keyword `activism' appears both in the politics of research and social movements; however, research on social movements focus on the activity as a research object: addressing `collectives' and `discourses', while politics of research has an inward looking perspective -- best illustrated with the keyword `hci education.'

Our category social issues includes the keyword `social movement', which is a separate category in our analysis.
This raises concern as to whether these two categories are truly separate or related to the same phenomena.
Work on social movements appeared more an external study of the phenomena while social issues often applied a critical or action-oriented approach.
To illustrate this difference, we categorised \citet{10.1145s2556288.2557100} as a social movement paper.
They highlight their contribution as ``\emph{present insights from an empirical analysis of data from an emergent social movement primarily located on a Facebook page to contribute understanding of the conduct of everyday politics in social media and through this open up research agendas for HCI}''.
In contrast, \citet{10.1145s3025453.3025490} serves as an example of the social issues category, and state their objective is ``\emph{developing a vision for a `Solidarity HCI' committed to designing to support personal social and institutional transformation through processes of agonistic pluralism and contestation where the aims and objectives of the solidary economy are continuously re-formulated and put into practice.}''
The former highlights more empirical approach on this, while the latter seeks to engage and address a social issue, more aligned with action-orientation.

We acknowledge that the keywords for motivating context or background-category are generic.
However, so are the use of words politics and democracy in the literature.
Papers in this category have been found to speak of ``\emph{politically and ethically sensitive materials}'' \citep{10.1145s2556288.2557196}, ``\emph{socio-political topology of the lived environment}'' \citep{10.1145s2858036.2858472} or to state that online venues facilitate discussions on ``\emph{topics ranging from political arguments to group coordination}'' \citep{10.1145s2998181.2998235}.
Authors may justify their work by stating that ``\emph{public, parliamentary and television debates are commonplace in modern democracies}'' but develop a tool for ``\emph{synchronous collaborative discussion of videos based on argumentation graphs that link quotes of the video, opinions, questions, and external evidence} \citep{10.1145s3317697.3323358}'' -- a contribution which is more distant from democracy than the initial sentence suggests.
Scholars have improved ``\emph{focus on socio-technical systems by taking seriously socio-political and socio-economic processes}'' \citep{10.1145s2145204.2145220} or highlighted that technological development is not ``\emph{independent of social, political or economic forces}'' \citep{10.1145s2675133.2675195}.
These are all legitimate uses of \texttt{politics} and \texttt{democracy}, but it does not appear to have a specific meaning: politics and democracy are part of society, and therefore they motivate our research aims, factor into the research findings, or suggest that there are political implications for new technologies.
At the same time, these uses do not provide a detailed purpose for invoking these words.
Rather, \texttt{politics} and \texttt{democracy} motivate or provide background for the scholarly contribution.
Therefore, while the category is less coherent in substance or contribution, the common denominator is a disconnect from politics or democracy beyond a brief mention.

We similarly categorised works related to academic community to a separate group, containing workshops, special interest groups, and doctoral consortium.
While such events could explicitly focus on some category above, at the same time they serve primarily for the purpose of establishing and maintaining academic communities.
Thus, they are different from categories where the goal is to disseminate novel scientific insights.

\subsection{Computational triangulation}
\label{computational_validation}

Computational analysis of the material supports the categories found during manual classification work.
As expected from previous research \citep{Baumer2017a}, topic model algorithm discovered more detailed topics than our thematic review process:
in total it identified  69 individual topics (see in Appendix~\ref{lda_details}, Figure~\ref{fig:stm_democracy}).
Topics correspond to to sets of words which are co-occure more commonly in the sets of documents:
for example in topic 52 identified co-occurances of (stemmed) words polit, social, use, twitter, activ, hashtag, media in some set of documents.
For us, this topic seems to correspond to the thematic category of user-generated content, as social media platforms and their use are clearly discussed.
There were several other topics -- 10, 11, 15 and 39 -- which seem to correspond to this topic, with characterising words such as populist, comment, hashtag, and contagion. 
We expand the diagnostics of the topic models in the Appendix~\ref{lda_details}.

Naturally, this analysis approach is crude.
For example, knowing that words co-occured in abstract does not tell how they were used, and more critically, it does not answer our semantic question.
Rather, as a tool for triangulation, we use this analysis to examine any mismatches between our manual approach and these outputs.

We do not identify mismatches which would require us to revisit our classification work from Stages 1A and 1B.
Each category had at least one corresponding topic, and many had more than one.
Most importantly, we did not observe new emergent themes from the topic-model approach.
There were in total 14 topics which we could not map into our manual classification.
These could be categories that our manual efforts did not observe.
Nonetheless, topic models often produce topics that are overly generic in the context of the research question.
We examined each of these 14 topics, but observed that they mostly consisted of generic scholarly words: user, design, system, technology, hci, research, etc.
Some topics focused on specific implementation technologies (topic 5: print, self-fold, model; topic: 24 print, thermoplast) or methods (topics 3: experi, particip, base, technolog, develop) but these do not speak about papers' relationship with democracy or politics.
Given that mismatches between these analysis approaches can be accounted on, we consider that this triangulation effort corroborates the manual classification work.

\subsection{Observations}
\label{sec:classification_discussions}

Our first observation is a high amount of content on academic community activities: for the keyword \texttt{democracy} this theme contains 49\% of all content and for \texttt{politics} 44\%.
Academic community activities include content focused on awareness-raising (such as panels and special interest groups) and early-stage research ideas (workshops, late-breaking work, posters, and other content that is not part of the main technical conference programme).
This suggests that the community is \emph{interested} in addressing politics and democracy, but this work does not follow through in full papers: for example, authors may have changed their framing away from these keywords and generic framings in their follow-up submissions.
To answer the semantic question, we focus on the remaining themes; they consist of fully developed conceptualisations.

\begin{table}

  \caption{The number of papers in each class.}
  \label{tab:rq2counts}

    \centering
    \definecolor{color1}{HTML}{CCBA51}
\definecolor{color2}{HTML}{4E6E3F}
\definecolor{color3}{HTML}{813BD7}
\definecolor{color4}{HTML}{BD4A6E}
\definecolor{color5}{HTML}{CE5BBF}
\definecolor{color6}{HTML}{845131}
\definecolor{color7}{HTML}{85D0B3}
\definecolor{color8}{HTML}{7F96C3}
\definecolor{color9}{HTML}{CEACA5}
\definecolor{color10}{HTML}{7BD353}
\definecolor{color11}{HTML}{D55B36}
\definecolor{color12}{HTML}{5D4599}
\definecolor{color13}{HTML}{42313F}

\begin{tabular}{llrr}
\toprule
 ~ & & \texttt{democracy} & \texttt{politics} \\
\midrule
\cellcolor{color1} & User-generated content &   8 &  38 \\
\cellcolor{color2} & Media consumption &   1 &  17 \\
\cellcolor{color3} & Public participation &  21 &  10 \\
\cellcolor{color4} & Social movements &   0 &   7 \\
\cellcolor{color5} & Social issues &   3 &  16 \\
\cellcolor{color6} & Public services &   2 &   4 \\
\cellcolor{color7} & Organization and workplace &   6 &   9 \\
\cellcolor{color8} & Politics of research &   1 &  19 \\
\cellcolor{color9} & Technology democratization &  21 &   0 \\
\cellcolor{color10} & Design methods &   5 &  30 \\
\cellcolor{color11} & Data analysis methods &   1 &  16 \\
\cellcolor{color12} & Motivating context or background &   9 &  46 \\
\cellcolor{color13} & Academic community activities &  74 & 166 \\
\bottomrule
\end{tabular}


\end{table}

We observe that the largest contributions to \texttt{politics} come from studies of media, both user-generated and media consumption themes.
Together, these two themes include 15\% of the material (or: 26\% when the academic community activities are removed).
This already highlights an answer to our semantic question: political communication is a subfield of (academic) political science and is a core context where human--computer interaction scholars connect technology and politics.
However, the politics that takes place in the media do not lead to a widening discussion of \texttt{democracy}: this consists of only 6\% of the corpus.
When used, word \texttt{democracy} seemed to serve as a way to discuss higher level concerns, such as ``\emph{We discuss the implications of these gender differences for democratic discourse and suggest ways to increase gender parity.}'' \citep{10.1145s3173574.3173907}

The word \texttt{politics} loads the theme `Motivating context and background' with 12\% of the corpus.
As we illustrated above, the theme collected ambigitious references to \texttt{politics}.
Politics and democracy were used to account for various factors related to real-world factors, but not to explain them further.
The phrase `political' was used in conjunction with other ambiguous terms, such as \emph{socio-political} or \emph{social, political or economic} factors.
Alternatively, for both \texttt{politics} and \texttt{democracy} we observed that the word was used to motivate technology development or deployment projects, albeit with a quick shift to a stronger technology development focus and contribution of the work.
The prominence of this theme suggests that politics and democracy belong to the jargon used by human--computer interaction due to their existence in society.
However, it seems that scholars were detached from a wider theoretic discussion:
they are somewhat vague placeholders for various unspecified phenomena (i.e., what is not a socio-political factor, which allows authors to acknowledge them but not focus on them in detail.
Similarly, when politics or democracy were used to motivate the research goal, the detailed connection between the contribution and politics or democracy is left unfocused.

The thematic group of civic society is driven by word \texttt{democracy} with 16\% of papers, but only 9\% of \texttt{politics}.
Similar to media themes, civic society themes seemed to be coherent in what they highlighted:
human--computer interactions identify politics and democracy in society.
Unlike in the `Motivating context and background' theme, these papers further elaborated what democracy and politics mean, and expanded these via concrete examples; for example:
``\emph{Strong representative democracies rely on educated, informed, and active citizenry to provide oversight of the government. We present Connect 2 Congress (C2C) [- -]}'' \citep{10.1145s1979742.1979852}.
The gist of this group is understanding politics and democracy in the context of the actions of individuals within structures:
democratic decision-making focuses on how individuals make decisions,
social movements focus on how individuals establish new structures to influence policy making,
and the social issues group ultimately focuses on examining established structures critically and even turn into activism.
Overall, as a semantic use it seems that politics and democracy in this thematic group involve citizens, while media themes their role seems to lie in mediating politics.

Interestingly, the word \texttt{democracy} is clearly more prominent in public participation, while the word \texttt{politics} is used more often in the context of social movements movements or with social issues.
This seems to mismatch some opportunities to wider discussion.
For example, in political sciences, social movements are seen as essential to democracy, but in our corpus, it appears that human--computer interaction scholars have not made such a connection.

Furthermore, the differences between the themes of social movements and social issues express differences in knowledge interests.
As we mentioned above, work on social movements focuses on works that describe and interpret civic activism.
To broaden the example set to illustrate this focus,
authors who belong to social movement state that they conduct an
``\emph{empirical analysis of data from an emergent social movement}'' \citep{10.1145s2556288.2557100} or that they
``\emph{analyze practices of political activists in a Palestinian village located in the West Bank.}'' \citep{10.1145s2556288.2557196}.
In contrast, work on social issues focus more directly on critiquing and challenging society, or proposing alternative social and technical arrangements for society.
Instead of highlight the work as analysis, these authors state they
``\emph{contribute to the socio-political designed innovation of solidarity movements}'' \citep{10.1145s2957276.2997020}
or that they report
``\emph{opportunities and challenges for supporting the development of local food networks with communities}'' \citep{10.1145s2675133.2675195}.
In a more abstract sense, some scholars may contribute more to descriptive and empirical research, while others engage more directly in normative questions.
For our semantic question, a difference in knowledge interests is essential:
while authors speak of politics, they may perceive these words to have different commitments in \emph{what} kind of work is expected to take place.

We also observe significant work involving discussion of research methods: in total 12\% of \texttt{politics} and 4\% of \texttt{democracy} papers.
In the data analysis method, the focus is on data that is somehow political: for example a political event can be used to gather data for development of machine learning methods \citep{10.1145s2645710.2645734}.
Therefore, politics is invoked in an instrumental sense: the (political) data serves as an illustrator of (political) data analysis.
This theme is similar to that of `Motivating context or background', but is more coherent in its core contributions.
Design methods instead are built to examine the politics of how design is conducted: politics serves as a word to reflect, for example, power relationships.

We also observe the phrase technology democratisation to drive the use of \texttt{democracy} (14\% of the corpus), which is a specific way to speak of rendering technology more accessible to wider population:
works sought to ``\emph{democratizes family access to technology}'' \citep{10.1145s3313831.3376344} or ``\emph{democratized access to data analytics}'' \citep{10.1145s1753846.1753872}.
Scholars also need to speak of normative perspectives and power conflicts related to research activity itself (5\% of \texttt{politics}).
Again, in this case politics is invoked to speak of a broad set of phenomena embedded into any human activity, but this group is specific as researchers are part of politics.
In the context of use, politics is invoked in a similar meaning, although the actors are not researchers but other people.
However, the topics are not so broad as to engage with the structures of society (on the civic society theme).

Overall, the thematic analysis extracts different dimensions of use for democracy and politics.
Some scholars use these words works where they seek to describe society, while work on the themes of social issues and democratisation had a more normative stance, asking how society could be improved.
Some themes, such as `Motivating context or background', used politics or democracy in an ambiguous sense to acknowledge that their work took place in a social context and used the word `politics' to cluster various social phenomena and ensure that they were addressed, but not fully opened up and explained.
This can be contrasted to themes such as `Media consumption', `Public participation' and `Social movements', which more extensively connected the research to society and political systems.
Therefore, as expected, human--computer interaction is semantically diverse when it uses \texttt{politics} and \texttt{democracy}, which creates a risk of speaking past one another.
For example, if politics is used to highlight reflection and the need for change (like in `Social issues' and `Design methods'), is this meaning lost for people who examine politics as an attribute of data points (`Data analysis method')?
What if some use it to account for various different phenomena to acknowledge that they understand their research in a wider socio-political context, while others use politics to engage with a specific phenomenon of the society and illustrate it?

\section{RQ2: Temporal developments in the themes}
\label{sec:rq2}

Table~\ref{tab:rq2counts} shows the total number of papers in each category over time, separated across the two corpuses.
We again observe that the corpuses increase in absolute volume, but as we observed with Table~\ref{tab:years}, this is attributable to the overall growth of SIGCHI publication volumes.
We observe that prior to the 21st century, politics and democracy were used in conjunction to fewer themes.
For both words, prior research referred mostly to organisations and workplaces: for politics, three papers focused on research methods were published before the 1990s, originating from a single research project developing interfaces for data analysis with political scientists.
Similarly, the role of academic community building and using these words to motivate and contextualise results have increased over the years, beginning from the early 2000s.
The increase has been more rapid in the academic community than for the motivational and contextual themes.

\begin{figure}
	\begin{subfigure}[t]{0.45\textwidth}
		\includegraphics[width=\textwidth]{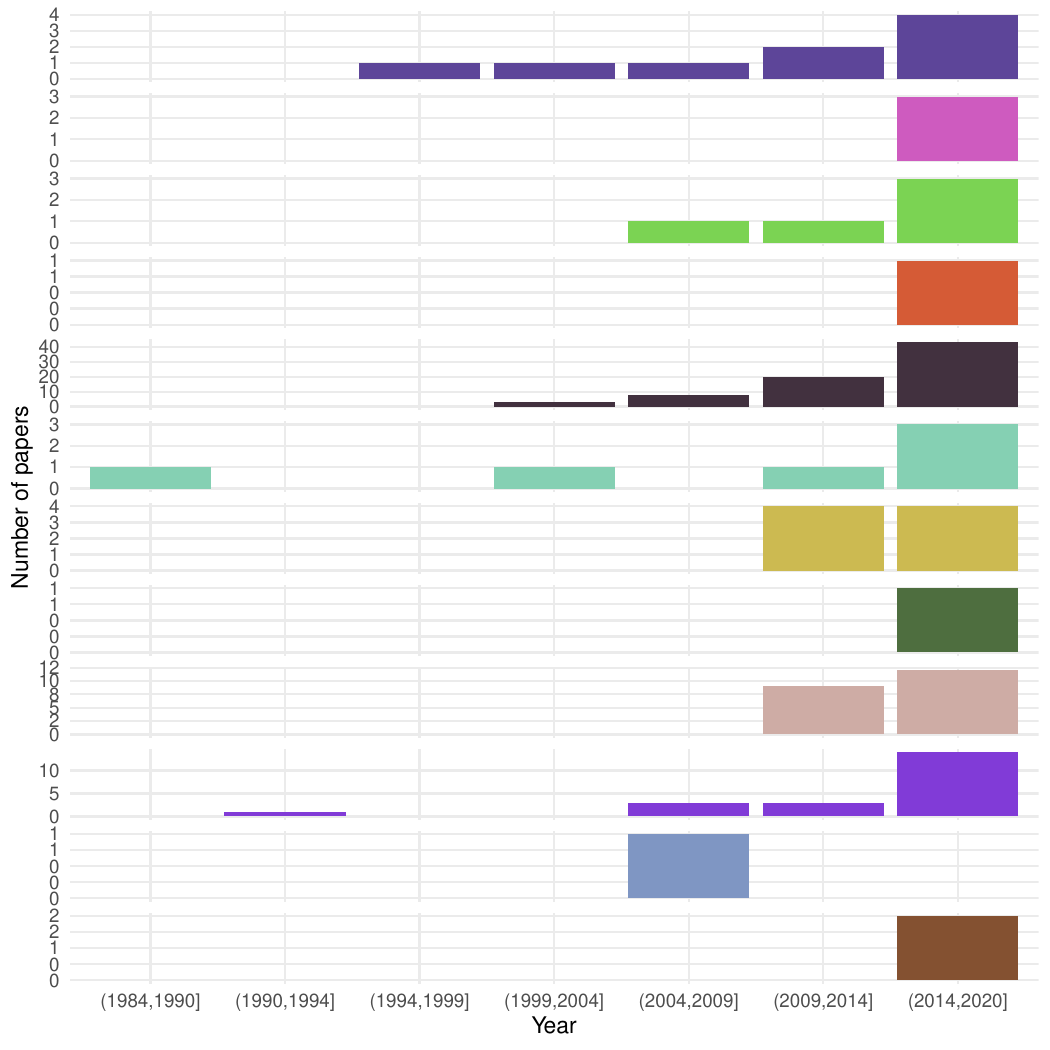}
		\caption{Timeline of categories in \texttt{democracy} corpus.}
		\label{fig:rq2timeline_democracy}
	\end{subfigure}
	~
	\begin{subfigure}[t]{0.45\textwidth}
		\includegraphics[width=\textwidth]{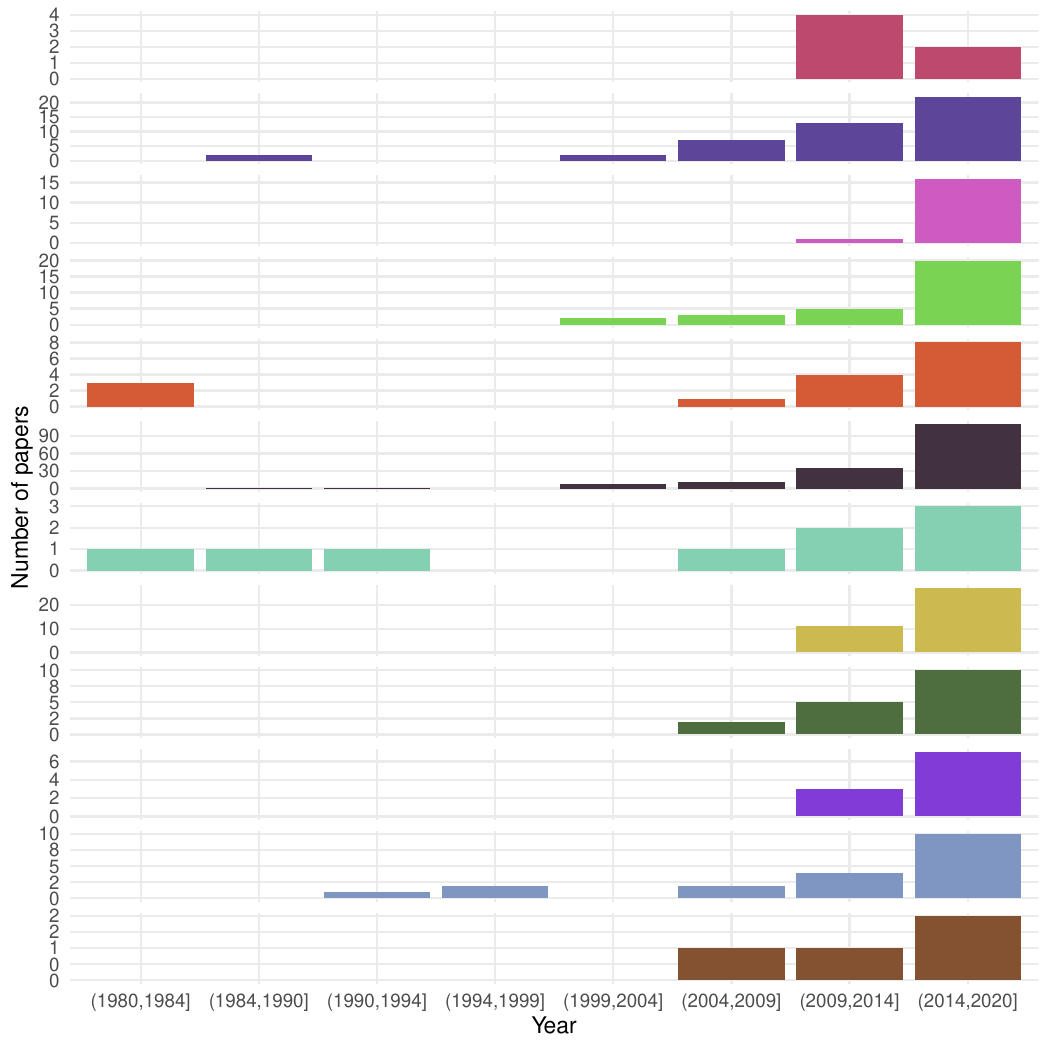}
		\caption{Timeline of categories in \texttt{politics} corpus.}
		\label{fig:rq2timeline_complete}
	\end{subfigure}

\definecolor{color1}{HTML}{CCBA51}
\definecolor{color2}{HTML}{4E6E3F}
\definecolor{color3}{HTML}{813BD7}
\definecolor{color4}{HTML}{BD4A6E}
\definecolor{color5}{HTML}{CE5BBF}
\definecolor{color6}{HTML}{845131}
\definecolor{color7}{HTML}{85D0B3}
\definecolor{color8}{HTML}{7F96C3}
\definecolor{color9}{HTML}{CEACA5}
\definecolor{color10}{HTML}{7BD353}
\definecolor{color11}{HTML}{D55B36}
\definecolor{color12}{HTML}{5D4599}
\definecolor{color13}{HTML}{42313F}
 
	\caption{Temporal analysis of publications.
 Colours same as in Table~\ref{tab:rq2counts}:
 \colorbox{color1}{user-generated content},
 \colorbox{color2}{media consumption},
 \colorbox{color3}{public participation},
 \colorbox{color4}{social movements},
 \colorbox{color5}{social issues},
 \colorbox{color6}{public services},
 \colorbox{color7}{organization and workplace},
 \colorbox{color8}{politics of research},
 \colorbox{color9}{technology democratization},
 \colorbox{color10}{design methods},
 \colorbox{color11}{data analysis methods},
 \colorbox{color12}{motivating context or background},
 \colorbox{color13}{academic community activities}.
 }
 
\end{figure}

Temporal analysis also shows how new connotations have been connected to politics and democracy over time.
In the period 2000--2010 we also observe the appearance of new themes; research on civic engagement and public services is connected to these words and the politics and democracy of design-based research methods.
In 2010--2015 discussion of technology democratisation began to take place and the politics and democracy of the media sector was examined.
In the media theme group, we see a difference in order with these words.
For democracy, user-generated content appears first and is followed by studies of media consumption, but for politics the order is reversed: media consumption is identified first and only then is user-generated content observed.

The clearest shift in research is visible with the theme of social issues, which clearly emerged rapidly during the 2015--2020 period, increasing 15-fold from the prior year in `democracy' (from one publication to 15) and appearing for the first time in the democracy word.
Critical and activist-oriented research on civic society is increasing; however, we observe that at the same time the absolute publication volume of more descriptive and observational research on social movements has seemed to decrease.
This is the only theme in which we observe a decrease in absolute publication volume during the 40 years of content analysed.

\section{Discussion}
\label{sec:discussion}

Over the past 40 years, human--computer interaction researchers have used the words \texttt{politics} and \texttt{democracy} in various ways.
Through a systematic literature review, we raise a semantic question: to what phenomena do we refer with these words and is it possible that these words have different connotations across human--computer interaction or lead us to speak across each other?
In the worst case, such differences can lead to decreased scientific productivity due to conceptual misunderstandings \citep{Oulasvirta2016,Abend2008}.
Indeed, our analysis of past publications in SIGCHI venues demonstrated a diversity in word use: we observed a total of 13 different categories, each showing a slightly different focus and potentially understanding of these words.
We also observed differences between \texttt{politics} and \texttt{democracy} and developments over the 40 years of analysis.
Next, we interprent these findings further by examining what our analysis reveals on human--computer interaction researchers' uses and thinking of politics and democracy.

\subsection{Orientations of politics and democracy}

Thirteen categories demonstrate various focuses scholars have on politics and democracy (see Table~\ref{tab:rq1result}).
To understand these better and identify semantic differences, we must contrast and compare these categories and the work conducted in them even more systematically.

Clearly, different categories consider a different core contribution for the work: some make a clear contribution to how technology and society relate to each other, while others clearly intended the core contribution to lie in more external societal perspectives.
For example, work on data analysis methods saw politics as a source of data, useful in illustrating the development and application of machine learning techniques to such data.
Their contribution is in the realm of data analysis, not on understanding or reflecting on society through such data.
However, similar work could also speak of society and its actions: work on user-generated content and media consumption could similarly use political data, but shed light on how politics is communicated.
This may partly explain some of the overlap in keywords across papers in these categories.
This difference may already manifest itself in a scholarly discussion:
data scientists engaged in politics and social media may actually seek to exploit pre-existing categorisation,
while a communication scholar engaged in politics and social media may expect a paper to have empirical implications as well.
While both scholars can say ``my work focuses on politics and social media,'' their scholarly outputs may look completely different: for the former, politics is of \emph{instrumental value}, but for the latter it is \emph{an object} of their research, both legitimate uses for words democracy and politics.

Our analysis indicates human--computer interaction scholars seemed to use the word politics (and democracy) to address phenomena at different levels.
The civic society and media categories appear to have a clear connection to society: the research focuses on actions that take place in public and impact a large group of people; politics and democracy appear to exist at the \textit{macro level.}
This can be contrasted to scholars who speak of politics in the context of organisations and workplaces: organisational politics rarely refers to societal-scale perspectives of politics.
Rather, organisational politics is \textit{mezzo-level} politics: power conflicts (which are essential to politics) emerge in organisations and information technology cannot be ignored in relation to them.
Similarly, work on design process and its political nature does speak of politics as an element of group interactions.
Papers on motivating context or background similarly address politics on a different level than the macro or mezzo-levels above: these papers acknowledge that political factors (often combined with other high-level concepts, such as social or economic) relate to their findings, but this insight is left unelaborated.
It appears that, in this category, politics is not seen as macro- or mezzo-level but rather \textit{out-of-reach} of more detailed analysis, but still requires acknowledgement.
For the semantic question, these different analysis levels highlight the complexity of both politics and democracy as words.
Three different levels of perspectives are used, but differences are not explicated, potentially leading into semantic confusion.

Third, the categorical difference between social movements and social issues shows how researchers' positions on politics vary within the body of research.
Scholars who used the word politics in conjunction to social movements appeared to approach the phenomena by observing them, while scholars engaged in social issues had a more normative and activist approach to technology research and development.
For decades, political and social scientists have discussed the tension between \textit{observational} and \textit{normative} research orientations:
is it the task of scholars to make politics, or to focus on describing and studying it as an object, with minimal involvement with it?
The discussion still continues within political science \citep[e.g.,][]{easton69,Marsh2004,Gerring2006a}.
Again, there is no right or wrong: both kinds of research are needed.
Rather, careful consideration should be put to ensure a plurality of researcher positions and ensure that single words are not loaded with normative stances or assumptions regarding correct ways of conducting research.
However, a clear semantic difference exists between works where politics is seen as something to be acted on and influenced rather than merely observed.

These three orientations demonstrate the inherent richness of research on politics and democracy: categories reveal differences in intended contributions, perspectives where politics lies, and even differences regarding researchers' positions.
While not dangerous, these different uses may create expectations across the community, which may require clarification.
For example, if a paper is deemed \emph{too political}, what does it mean:
does the paper take an activist stance, whereas the reader expected a more observational analysis
or
does it engage so much with political institutions that it is no longer about human--computer interaction but political science?
Similarly, what would a \emph{more democratic} paper look like:
greater engagement with society,
wide-ranging remarks on democratic implications of the technology,
or something else?
More broadly, while we have diverse ways of using words \texttt{democracy} and \texttt{politics}, is our focus balanced across the various perspectives?
To examine these questions further, we next examine temporal changes and differences between these two keywords and examine the semantic question through these lenses.

\subsection{Temporal change and expanding meanings of politics and democracy}

We also show an increasing expansion of both terms to novel categories over the 40 years of analysis.
Originally, both stemmed from attempts to understand small groups, such as organisations and workplaces.
Politics had in addition a few papers focused on how political scientists used computers in their data analysis and what design needs were emerging for supporting political data analysis.
This is in clear contrast to the late 2010s, when active research was conducted in 13 different categories.

Semantically these categories indicate three shifts away from organisational democracy and politics and the expansion of research under the umbrella.
The first shift, which began in the 1990s, started to acknowledge the existence of politics in technology research, indicated by the increase in the `Motivating context or background' category.
While this shift acknowledges politics and democracy, this category may indicate that these topics were seen as beyond the scholars' reach.
A second shift moves the politics into reach by examining politics and democracy in the context of media systems and civic society.
The most recent shift, taking place in late 2010s, brought up a critical stance towards politics and democracy, showcased by the `Social issues' category.
Therefore, the temporal analysis reveals the changes through which orientations seem to gain attention in our field.

These orientation changes seem surprising, as there have not been massive changes in how society perceives democracy and politics over the past 40 years.
While we have observed the collapse of the Soviet regime, a still ongoing decrease in voting activity and the emergence of populist political positions, there is no end to politics in sight or a clear explanation for why orientations have changed.
In part, these changes mirror the rapid digitalisation of our society:
media, while a decades-old political science research topic, only became relevant to human--computer interaction when news organisations begun to publish content on the Web.
This explains the emergence of the `media consumption' category.
However, technology change alone does not explain these changes.
Political scientists have studied user-generated content since the era of MUDs, email lists and newsgroups \citep{resnick1997politics,Dahlberg2001,Papacharissi2004a}, but for human--computer interaction the phrasing of user-generated content in conjunction to democracy and politics emerged in the era of social media.
Similarly, technological change is nothing new: it has fueled the transformation of political parties for decades \citep{farrell00}.
In particular, more critical and activist stances on politics and democracy research have traditions that extend far before the human--computer interaction researchers' focus on the late 2010s.

These changes therefore cannot be explained only by changes in the political or technical environment.
Instead, these changes may reflect how the field -- like the wider technology sector -- has begun to appreciate politics and democracy and thus invoke these terms in their research.
The new orientations attached to politics and democracy indicate how the field's thinking of democracy and politics is evolving.
This indicates an ongoing semantic drift in the field to focus research into different areas of politics and democracy.

For such evolution, we must ask whether there are blind spots in our fields' current semantic focuses.
For example, according to \citet{Nelimarkka2019a}, human--computer interaction research on democratic decision-making has seemed to mostly ignore representative systems and focus on citizen-led forms of democracy.
He argues that this is a clear omission, which has also harmed the development of the field.
Semantic analysis and categories may help us to start asking the question: is there more to politics than what our analysis finds in the corpus?
For example, we observed decrease of work on social movements,
none of the works we examined engaged on the human body itself as a political,
nor examined if economic systems as democratic.
More critically, human--computer interaction reseaarchers work on topics such as elections and social justice,
but our analysis suggests these scholars are less likely to invoke words democracy and politics (and corresponding theories) to connect the work to wider context.

\subsection{Differences between politics and democracy}

Our analysis also highlights the various uses of words `politics' and `democracy' in human--computer interaction research.
Naturally, these terms do have diverse colloquial meanings as well, not to mention how these terms are used within political science.
Therefore, the differences as such are not striking.
Some differences also stem from specialised jargon: for example the phrase `democratisation' is specific to a particular research movement and includes the word \texttt{democratisation}, thus establishing this category in the democracy corpus, while the same focus is not present in the politics corpus.
However, differences also help us understand the semantic differences between these words and their uses.

For example, media-related research is much more strongly about politics than about democracy.
In other words, while human--computer interaction scholars have invoked politics in this context, we observe fewer papers at the level of title, abstract and keywords engaging with democracy in the context of media.
It therefore seems that while politics is widely acknowledged, the reflection may not follow through to understanding the system-level implications for democracy.
Democracy is invoked when speaking about public participation and decision-making procedures, a topic that is less often addressed from the perspective of politics.
Again, it seems that these differences stem from perspective bias: based on keywords, abstracts and titles, it does not seem that works examining public participation through technology examine the politics of public participation.
Often the ultimate aim of public participation is to allocate limited resources, such as city space, or make formal decisions -- they may be prone to power conflicts, which are at the core of politics.

Therefore, there are semantic difference between these words -- as they are different words, such differences are expected.
However, these differences also indicate perspective differences that may require more careful consideration within our community.
Should we ensure that more papers looking at political communication in mediated media environments engage not only with politics, but also, in their abstracts, draw a connection to democracy -- an inherent part of political communication?
Similarly, when seeking to advocate democratic decision-making or increasing access to technology, we should also be careful to identify whether we wish to speak about the politics regarding these themes.

Similarly, we observed a temporal difference within the media categories.
User-generated content first appeared in the democracy corpus, while media consumption was first observed in the politics corpus.
This further demonstrates how these terms draw on different ideas and goals, even while later on both categories are studied from both perspectives.

\subsection{Approaching democracy and politics in the future -- towards a research agenda}

We have this far focused on a semantic question: what meanings and contexts are attributed to words democracy and politics.
However, our findings beg the question: how could human--computer interaction scholars approach democracy and politics in the future and improve our understanding of this vital and increasingly popular topic.

First, we observed that on the abstract and titles, these two terms attributed different perspectives: for example, public participation was more about \texttt{democracy} while social movements were about \texttt{politics}.
From political science perspective, such division of use seems peculiar: public participation is political -- it has interest conflicts among people which need to be resolved -- and social movements around political issues are essential for functioning democracy, i.e. the system where people’s preferences ought to share state-level decision-making.
Therefore, co-use of these terms even on abstracts might open the research subjects from a new light and help scholars to examine more deeply their connections.

Second, we suggested that there are two different researcher position: observational and normative; as highlighted above, such a division exists also in political science \citep[e.g.,][]{easton69,Marsh2004,Gerring2006a}.
Both positions provide valuable insights regarding the society, but analytically their co-existence open new questions.
Overall, a question meriting further discussion is how to share scholarly attention across these two positions.
In addition, should normative works defend their stance when, i.e. if scholars seeks to support a social cause, what kind of elaborations are expected from them?
And when judging such works, are we assumed to distance ourselves from the case and its societal goals or should we also evaluate them -- and if those are evaluated, how we maintain a pluralist perspective regarding the society?
As polarisation is increasing in societies and global tensions are on the rise, differences across value systems are becoming more prominent. 

Last, we observed different levels: democracy and politics were seen out-of-reach, macro- or mezzo-levels.
The latter two focus on societies and groups, but should human--computer interaction research also focus on the individual, that is, the  micro-level?
For example, individuals' political believes and ideologies and their connection to interactive system design, user experience etc. would open the individual as a political entity.
This level highlights an additional focus for scholarship.

\subsection{Limitations and future work}

Most importantly, our literature selection narrowed the focus to papers discussing politics and democracy \emph{explicitly} and to SIGCHI-sponsored publication venues.
This is a suitable approach to examining the semantic question, but it means that our study is not about all things political or democratic (recall Figure~\ref{fig:semanticliteraturesearch}),
Different methods of selecting a corpus for analysis may afford additional insights.
That direction can be considered to offer natural room for expanding on our work.
For example, efforts could be directed to examining literature outside SIGCHI venues and contrast if there these words are offered different semantic meanings across the board of scholars.
However, careful effort is required to choose how to scope this review, as human--computer interaction itself is not clear-cut.
We believe that adding venues such as \emph{Transactions of Human--Computer Interaction} or \emph{European Journal of Computer--Supported Collaborative Work} might not be objected.
However, where we define the boundaries: should we also examine more human--computer interaction -oriented works in information systems or design studies?
Should some political and social science works be included if they focus on developing novel information systems and experimentally studying them?
While the focus on ACM only venues is limited, it captures many of the highest impact venues.
In the context of our research question, it is possible that the results are leaned towards North-American perspective of politics and democracy,
however the results do not seem to focus on particular cultural or institutional settings.
For example, studies on workplace and democracy may be more prominent in the European computer--supported collaborative work scholarship, but this semantic meaning would have been given.
Therefore, we believe that biases in RQ1 are limited, the temporal analysis on RQ2 is more sensitive to our focus SIGCHI.
Similarly, the semantic question and answers to them could be different if words like governance, goverment, or power were used: other keyword sets or literature-selection approaches could be used to increase our knowledge of this domain further.

Another area for future work is deeper engagement with the literature.
We opted to conduct theme-based analysis only.
While this is an accepted method for a literature review \citep{Peng2012,Suominen2016,Griffiths2004,Bossen2017},
detail-level examination -- looking at, for instance, a single category in depth -- could encourage more concrete proposals for research.
Ideally, detailed analysis could show how political science may benefit from the outputs of human--computer interaction research and aid in setting a more explicit research agenda for human--computer interaction research.
Secondly, we did not attempt to study the intellectual and ideological roots of this literature.
For example, citation network analysis could yield insights in this regard
and thereby create a foundation for
identifying the core concepts used in the relevant scholarship,
or even
for critical study of how the core concepts are used and whether the use corresponds to the domain-specific understanding of them, along the lines of what \citet{Marshall2017} did for ``performance'' and \citet{Schmidt2016a} for ``awareness.''
Lastly, a detailed analysis of funding instruments acknowledged and collaboration networks could improve our understanding of why we have observed particular temporal trends.

Finally, we must reiterate that classifications cannot be value-neutral.
In our analysis, we conducted extensive characterisation and classification of politics and democracy, and these analysis steps should be seen as political (in the politics-as-social-order sense) \citep{10.2307/44483288}.
As we have discussed above, we are supportive of more inclusive definitions for politics,
although our work had to apply a narrower scope to limit the selection criteria.
Future research could engage explicitly with anarchist, feminist, and post-colonialist literature in the human--computer interaction discipline (and its counterparts in political science) \citep{Keyes2019,bardzell2010feminist,bardzell2011towards}.
One should also remember, however, that this work and the classifications therein are not intended to define \emph{the} meaning of politics and democracy in relation to human--computer interaction.
Rather, our work provides \emph{a} meaning, and future work is welcome.
Continued investigation and reflection surrounding politics can only benefit our discipline, other fields, and society as a whole.

\subsection{Concluding discussion: What about the semantic question?}

Our review began with a \emph{semantic question} \citep{Abend2008}: when do human--computer interaction scholars speak about \texttt{politics} or \texttt{democracy}?
We conducted a systematic thematic and temporal analysis of both terms in the ACM Digital Library and identified a total of 13 different categories of use, aligned with six more wider topics.
These categories open up \emph{orientations}: contribution focus, level of studied phenomena  -- mezzo, macro or out-of-reach -- and researchers' position.
These orientations demonstrate the variety of meanings that researchers can attribute to work when describing it with words like \texttt{politics}, \texttt{political}, \texttt{democratic} etc.
At the same time, that variety means that we must be careful to acknowledge pluralism in relation to such terms.
To aid communication across different orientations, authors could pinpoint even more explicitly what is democratic or political for them.
Furthermore, scholars working on more civic-oriented phenomena might consider how to better integrate both concepts into their work: they open up different perspectives that may help us better engage, for example, with the implications or group dynamics that relate to our research.

We also demonstrate that the semantic meaning seems to be expanding.
Over time, new levels of phenomena have changed what human--computer interaction researchers study.
Most recently, we could detect a turn to more action research-oriented styles of doing work, where politics is \emph{engaged}, not only observed.
Such expansions provide richer phenomena for our scholars and demonstrate our communities' understanding of how politics and democracy relate to technology.
This seems to relate partly to technological development, but this does not sufficiently explain the expansion.
Rather, it appears that our community is drawing on new ideas and perspectives to help itself work with politics and democracy.
We hope that this literature review and semantic research helps to systematise this exchange of ideas.

Overall, our study shows that, while the work is increasingly diverse and accounts for different orientations, the overall volume of publications is still relatively low -- even today, politics accounts for about 1\% of total publications, and democracy is addressed in less than half of these.
At the same time, the largest groups in both terms relate to building academic community: special interest groups, panels, workshops, poster presentations, etc.
However, our examples show that, semantically, politics and democracy is not only about Politics, that is, governments and society-level actions.
They can also be seen in organisations, service delivery and emerging during the design.
While further engagement may blur the semantic coherence of the concept even more, how might the lens of politics and democracy -- when further elaborated -- help our community to increase its social relevance?

\section*{Acknowledgments}

Blind for review.


\bibliographystyle{ACM-Reference-Format}
\bibliography{mattiMendeley,extra}

\newpage

\appendix

\section{Computational triangulation of classification schema}
\label{lda_details}

\begin{figure}[!b]
	\includegraphics[width=.95\columnwidth]{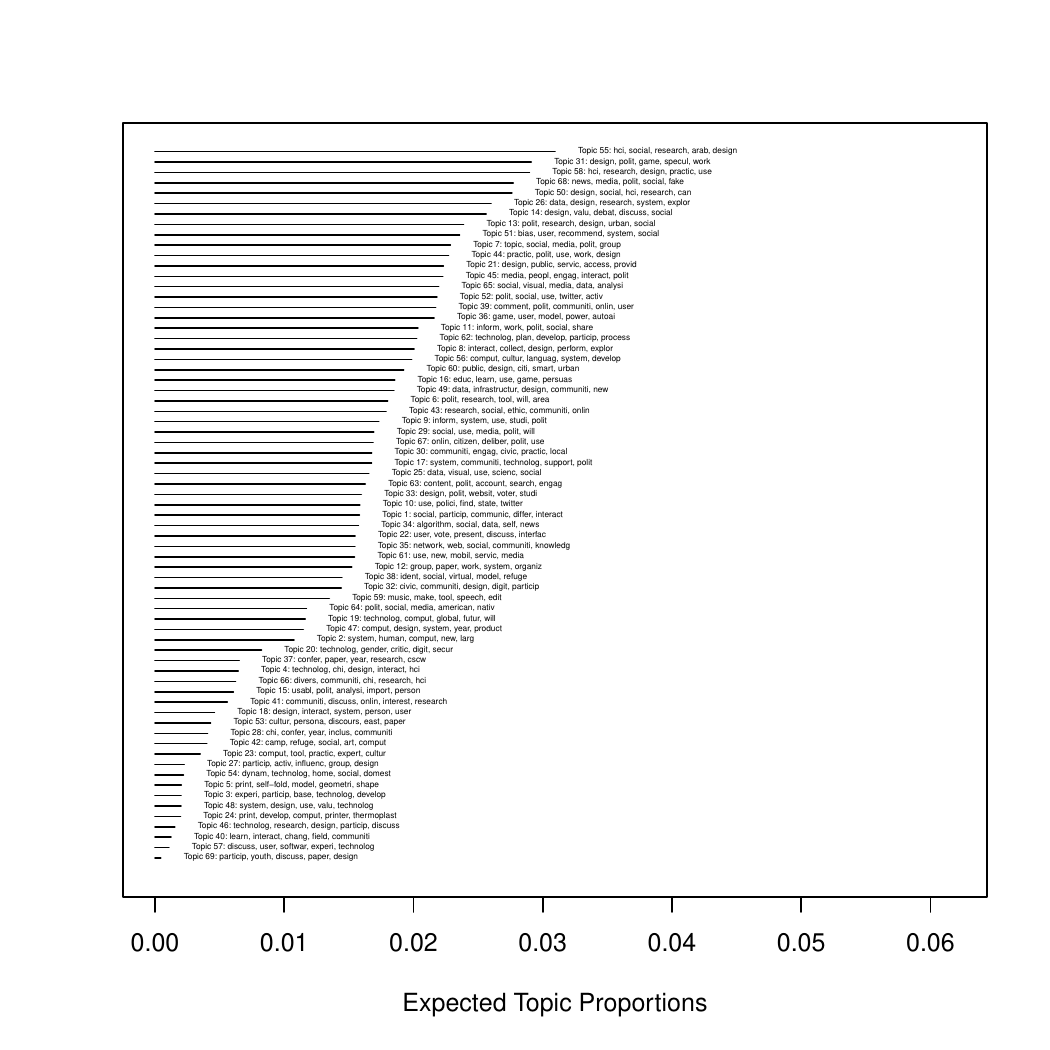}
	\caption{Topics and the estimated proportions for their representation in the abstracts. Enlarge view for details.}
	\label{fig:stm_democracy}
\end{figure}

We conducted an additional triangulation of the classification framework by executing a topic model on the abstracts.
We used structural topic models (STM) and optimised the number of topics to produce the best-fit model.
The best-fit model had 69 topics, illustrated in Figure~\ref{fig:stm_democracy}.
As \citet{Baumer2017a} have pointed out, topic models often produce a low-level classification of the data.
Therefore, we examined the output and, as necessary, merged topics into clusters to set the triangulation's component classifications at the same conceptual level.

We observed following clusters to emerge from the data, corresponding to our manual classification framework, presented by highest propability or highest lift keywords:

\begin{description}
    \item[User-generated media] Topic 10 (use, polici, find, state, twitter, differ, platform), Topic 11 (inform, work, polit, social, share, media, differ), Topic 15 (usabl, polit, analysi, import, person, polici, tweet), Topic 39 (comment, polit, communiti, onlin, user, aapi, social), Topic 52 (polit, social, use, twitter, activ, hashtag, media)
    \item[Media consumption] Topic 34 (algorithm, social, data, self, news, media, use), Topic 63 (content, polit, account, search, engag, politician, user), Topic 68 (news, media, polit, social, fake, bias, user).
    \item[Public participation] Topic 22 (user, vote, present, discuss, interfac, adapt, correct), Topic 29 (social, use, media, polit, will, interact, program), Topic 30 (communiti, engag, civic, practic, local, govern, form), Topic 32 (civic, communiti, design, digit, particip, live, analysi), Topic 33 (design, polit, websit, voter, studi, inform, work), Topic 45 (media, peopl, engag, interact, polit, tool, civic), Topic 67 (onlin, citizen, deliber, polit, use, platform, nudg), Topic 69 (particip, youth, discuss, paper, design, social, polit)
    \item[Social movements] Topic 27 (particip, activ, influenc, group, design, individu, benefit), Topic 35 (network, web, social, communiti, knowledg, polit, user), Topic 64 (nativ, nonprofit, american, advoc, elect, immigr, africa)
    \item[Social issues] Topic 17 (system, communiti, technolog, support, polit, design, activist), Topic 20 (technolog, gender, critic, digit, secur, wearabl, women), Topic 36 (autoai, playabl, autoaiviz, logic, monopoli, diseas, burden), Topic 38 ( ident, social, virtual, model, refuge, avatar, relat), Topic 40 (learn, interact, chang, field, communiti, program, challeng), Topic 42 (camp, refuge, social, art, comput, palestinian, exhibit), Topic 44 (practic, polit, use, work, design, studi, collabor) Topic 55 (hci, social, research, arab, design, solidar, work).
    \item[Public services] Topic 13 (polit, research, design, urban, social, citi, hci), Topic 21 (design, public, servic, access, provid, citizen, project ), Topic 26 (iot, thing, disast, tangibl, environment, internet, sens), Topic 60 (smart, citi, display, pollut, urban, public, street)
    \item[Organisations and workplace] Topic 1 (employe, cmc, voic, touch, workplac, agon, construct), Topic 2 (human, workspac, feedback, larg, extend, interfac, robot), Topic 16 (educ, persuas, worker, health, bodili, persuad, robot )
    \item[Design methods] Topic 8 (interact, collect, design, perform, explor, food, process), Topic 14 (design, valu, debat, discuss, social, media, argument), Topic 31 (specul, graphic, altern, object, game, privaci, research-cr), Topic 53 (persona, east, stereotyp, dimens, rubio, contradict, asian)
    \item[Data analysis methods] Topic 7 (topic, social, media, polit, group, discuss, user), Topic 51 (bias, user, recommend, system, social, predict, polit), Topic 65 (social, visual, media, data, analysi, inform, use).
    \item[Professional topics] Topic 9 (computer, closur, flood, curricula, religi, ubicomp, diet), Topic 19 (global, transnat, futur, scholar, mobil, profession, consid), Topic 25 (cut, flag, inter-univers, machine-read, energi, segment, verifi)
    \item[Technology democratization] Topic 54 (dynam, technolog, home, social, domest, participatori, need), Topic 59 (music, make, tool, speech, edit, democrat, diy)
    \item[Motivating context or background] Topic 56 (comput, cultur, languag, system, develop, interact, technolog)
    \item[Academic community activities] Topic 4 (profound, strike, implant, simpli, portabl, two-day), Topic 6 (area, european, wall, cycl, presidenti, panelist, current), Topic 12 (group, paper, work, system, organiz, confer, scienc), Topic 28 (chi, confer, year, inclus, communiti, person, stori), Topic 37 (confer, paper, year, research, cscw, will, process), Topic 50 (team, address, invit, facilit, ethic, participatori, workshop), Topic 62 (plan, sig, urban, citizen, anim, participatori, ictd),  Topic 66 (divers, chi, inclus, effort, award, true, subject)
\end{description}

As the list illustrates, some of the topic assignments to a single individual category are difficult, for example Topic 10 could consists of papers across various categories which study Twitter use and address the platforms.
Therefore, the topic could also belong to categories such as media consumption, social movements, or data analysis methods, depending on the contribution of the article.
Nonetheless, this should be considered as an indicator that these categories exist on the data, even while at such high level we may make mistakes on allocating topics to these categories.
Similarly, topic models do a poor job with the Academic community activities-category.
While it shows workshops, sigs, panels etc. do exist in the corpus, it does not capture the content them.
The substance of community activities is distributed to other categories.
This is different from the manual classification, where we choose to draw them together into a single category.
However, as we wish to highlight the significant portition of them in the whole dataset, this single-category classification for the content can be justified.

As is often the case with topic models, the analysis produced several topics that could be discarded from further analysis.
Given our focus on triangulation, these categories must be vetted to ensure they do not represent new categories missed from our analysis work.
We can identify some of these as generic terms uwhich could have been removed as stop words: topics 46 (technolog, research, design, particip, discuss, will), 57 (discuss, user, softwar, experi, technolog, use, particip), and 61 (use, new, mobil, servic, media, develop, polit) represent such generic words.
Similarly, there are stop-word like terms more specific to human-computer interaction research:
\begin{itemize}
\item scholars work with computers, systems and infrastructures -- as represented by topics 18 (design, person, interact, well, involv, system), 23 (comput, tool, practic, expert, cultur, product, artifact), 48 (system, design, use, valu, technolog, societi), 47 (comput, design, system, year, product) and 49 (data, infrastructur, design, communiti, new, work, develop)
\item they use spesific methods and data sources -- represented in topics 3 (experi, particip, base, technolog, develop), topic 41 (communiti, discuss, onlin, interest, research), topic 43 (ethic, signal, consent) and 58 (hci, research, design, practic, use, communiti)
\item some use unique technologies which are algorithmically identified as their own topics but do not address the relationship with democracy (topic 5: print, self-fold, model, topic 24: print, develop, comput, printer, thermoplast).
\end{itemize}

These account for topics not originally identified in our framework and gives a reasonable rational that they do not produce additional answers to our research questions.
Therefore, we do disregard these topics from further analysis and do not believe they challenge the manual classification work.

\newpage

\section{Details on classification work}
\label{classification_schema_2_details}


\begin{longtable}{p{.1\textwidth}p{.4\textwidth}p{.25\textwidth}p{.25\textwidth}}

    \label{tab:rq1result_details}
    \\ \toprule
Category & Examples & Keyword: \texttt{politics} & Keyword: \texttt{democracy} \\
\midrule

\endfirsthead

\toprule
Category & Examples &  Keyword: \texttt{politics} &  Keyword:  \texttt{democracy} \\
\midrule

\endhead

\midrule
\multicolumn{4}{l}{\emph{Continued on next page}} \\
\bottomrule
\endfoot

\bottomrule
\endlastfoot

\multicolumn{4}{l}{\textbf{Media}} \\
User generated content &
``A Characterization of \underline{Political} Communities on Reddit'' \citep{10.1145s3342220.3343662},
``Constructing the Visual Online \underline{Political} Self: An Analysis of Instagram Use by the Scottish Electorate'' \citep{10.1145s2858036.2858160},
``microblogging information diffusion activity during the 2011 Egyptian \underline{political} uprisings'' \citep{10.1145s1165387.275654},
``characterize users who adversarially interact with \underline{political} figures on Twitter [- -] in the two months leading up to the 2018 midterm elections'' \citep{10.1145s2631488.2631498},
and
``\underline{Political} Hashtags \& the Lost Art of \underline{Democratic} Discourse'' \citep{10.1145s3025453.3025543}, 

& \citepreview{10.1145s1165387.275654,10.1145s2145204.2145212,10.1145s2998181.2998194,10.1145s3342220.3343662,10.1145s3317697.3323352,10.1145s3313831.3376548,10.1145s3173574.3174210,10.1145s2858036.2858160,10.1145s2675133.2675163,10.1145s2856767.2856776,10.1145s2702123.2702291,10.1145s2702123.2702403,10.1145s3290605.3300836,10.1145s2998181.2998295,10.1145s3027063.3053185,10.1145s2998181.2998287,10.1145s2858036.2858423,10.1145s2858036.2858398,10.1145s2531602.2531719,10.1145s3342220.3343658,10.1145s3173574.3174207,10.1145s2441776.2441877,10.1145s2531602.2531676,10.1145s2702123.2702193,10.1145s2556288.2557197,10.1145s2675133.2675187,10.1145s2441776.2441935,10.1145s3313831.3376542,10.1145s2792838.2800190,10.1145s3025453.3025543,10.1145s3196709.3196764,10.1145s2531602.2531605,10.1145s2145204.2145213,10.1145s1958824.1958842,10.1145s3342220.3343657,10.1145s2531602.2531735,10.1145s2441776.2441876,10.1145s3372923.3404817,10.1145s3148330.3148336} & \citepreview{10.1145s2858036.2858160,10.1145s2449396.2449424,10.1145s2531602.2531719,10.1145s2531602.2531677,10.1145s2675133.2675134,10.1145s3313831.3376542,10.1145s3025453.3025543,10.1145s2531602.2531605} \\

Media consumption &
``Broadening Exposure to Socio-\underline{Political} Opinions via a Pushy Smart Home Device'' \citep{10.1145s3313831.3376774},
``Is a polarized society inevitable, where people choose to be exposed to only \underline{political} news and commentary that reinforces their existing viewpoints?'' \citet{10.1145s1978942.1979106} , and
``When one searches for \underline{political} candidates on Google, a panel composed of recent news stories, known as Top stories, is commonly shown at the top of the search results page.'' \citet{10.1145s286498.286538}

& \citepreview{10.1145s286498.286538,10.1145s1753326.1753541,10.1145s1978942.1979127,10.1145s3313831.3376774,10.1145s1978942.1979106,10.1145s3290605.3300300,10.1145s3290607.3299046,10.1145s3025453.3025833,10.1145s3281151.3281157,10.1145s1518701.1518772,10.1145s1753326.1753543,10.1145s2998181.2998321,10.1145s3025453.3025611,10.1145s3372923.3404794,10.1145s2441776.2441895,10.1145s3290605.3300820,10.1145s1518701.1518824,10.1145s3313831.3376232} & \citepreview{10.1145s3343413.3377975} \\

\multicolumn{4}{l}{\textbf{Civic society}} \\
Public participation &
``Factful, a web-based annotative article reading interface that enhances the article with fact-checking support and contextual budgetary information by processing open government data. [- -]'' \citep[keyword: deliberative \underline{democracy}]{10.1145s2702123.2702352},
``[- -] Information and Communications Technologies (ICTs) support forms of community activism that operate outside formal \underline{political} and institutional channels. We have done fieldwork with local housing justice activists in order to gain insight into the way ICTs play a role in complementing forms of civic engagement that challenge, rather than work with, institutional authority.[- -]'' \citepreview{10.1145s2675133.2675156},
``[- -] the outcome of a concerted effort to develop responsive and impactful direct \underline{democracy} platforms. We offer a sociotechnical genealogy of the process, informed by theory of deliberative \underline{democracy}'' \citep{10.1145s3290605.3300300}, and
``Strong representative \underline{democracies} rely on educated, informed, and active citizenry to provide oversight of the government. We present Connect 2 Congress (C2C), a novel, high temporal-resolution and interactive visualization of legislative behavior.'' \citep{10.1145s1979742.1979852}

&  \citepreview{10.1145s2317956.2317982,10.1145s2998181.2998291,10.1145s3270316.3271513,10.1145s3025453.3025996,10.1145s2675133.2675156,10.1145s3313831.3376603,10.1145s2598510.2598523,10.1145s3173574.3173901,10.1145s2145204.2145249,10.1145s3290605.3300517} & \citepreview{10.1145s1499224.1499273,10.1145s2998181.2998291,10.1145s1457199.1457229,10.1145s3025453.3025853,10.1145s1753846.1753872,10.1145s3313831.3376293,10.1145s3290605.3300247,10.1145s3027063.3053248,10.1145s2702123.2702352,10.1145s2675133.2675156,10.1145s3313831.3376603,10.1145s1385569.1385652,10.1145s3313831.3376158,10.1145s2675133.2675212,10.1145s2598510.2598523,10.1145s2858036.2858098,10.1145s3357236.3395531,10.1145s1125021.1125063,10.1145s3173574.3174148,10.1145s3334480.3382964,10.1145s2207676.2208594} \\

Social movements &
``[- -] In this paper we present insights from an empirical analysis of data from an emergent social movement primarily located on a Facebook page to contribute understanding of the conduct of everyday \underline{politics} in social media and through this open up research agendas for HCI. [- -] We outline possible research agendas in the field of everyday politics, which are sensitive to the everyday acts of resistance enclosed in the ordinary.'' \citep{10.1145s2556288.2557100},
``analyze practices of \underline{political} activists in a Palestinian village located in the West Bank.'' \citep{10.1145s2556288.2557196}, and
``Transgender people are marginalized, facing specific privacy concerns and high risk of online and offline harassment, discrimination, and violence. Participants frequently returned to themes of activism and prosocial behavior, such as protest organization, \underline{political} speech [- -]'' \citep{10.1145s1978942.1979290}

&\citepreview{10.1145s2556288.2557100,10.1145s2470654.2466262,10.1145s2531602.2531611,10.1145s2858036.2858153,10.1145s3313831.3376339,10.1145s2531602.2531664} & \\

Social issues &
``Sex workers' rights are human rights, and as such are an issue inherently based in social, criminal, and \underline{political} justice debates. As HCI continues to move towards feminist and social justice oriented research and design approaches, we argue that we need to take into consideration the difficulties faced by sex workers [- - ] We discuss their service provision and the ways in which HCI is uniquely positioned to be able respond to the needs of and to support sex work support services'' \citep{10.1145s3025453.3025615},
``Through embedded work with several solidarity structures in Greece, we have begun to understand the solidarity economy (SE) as an experiment in direct \underline{democracy} and self-organization.'' \citep{10.1145s3025453.3025490}, and
``Solidarity organizations in Europe are committed to building a more socially just society through a better configuration of \underline{democracy}, \underline{politics} and economy. In this paper, we describe our efforts to contribute to the socio-\underline{political} designed innovation of solidarity movements through the establishment of a research lab embedded in, and operating within, the solidarity economy.'' \citep{10.1145s2957276.2997020}
& \citepreview{10.1145s3209542.3210576,10.1145s2631488.2631498,10.1145s2702123.2702283,10.1145s2702123.2702157,10.1145s3290605.3300810,10.1145s2818048.2820027,10.1145s3322276.3322339,10.1145s3025453.3025490,10.1145s3173574.3174055,10.1145s3173574.3173646,10.1145s3025453.3025961,10.1145s2957276.2957311,10.1145s2858036.2858343,10.1145s2818048.2820026,10.1145s3025453.3025615,10.1145s3173574.3173880,10.1145s3173574.3173960} & \citep{10.1145s3173574.3173907,10.1145s3025453.3025490,10.1145s3173574.3174055} \\
\multicolumn{4}{l}{\textbf{Context of use}} \\

Public services &
``Digital technologies offer the possibility of community empowerment via the reconfiguration of public services. This potential relies on actively involved citizens engaging with decision makers to pursue civic goals. [- -] We offer a number of design considerations for future HCI research, focusing on how digital technology might be configured more appropriately to support campaigning around the \underline{politics} of mobility.'' \citep{10.1145s2858036.2858146},
``Social computing provides a new way for citizens to engage with their public service. Our research investigates how social computing might support citizens co-design their transit service. (Keyword: \underline{political} design)'' \citep{10.1145s1979742.1979835}, and
``[- -] We address this gap by engaging these elements in ongoing design research within Atlanta's Department of Immigrant Affairs. Our research inquiry with the department centered on developing a design intervention to improve the department's community engagement work. (Keyword: \underline{digital} democracy)''
& \citepreview{10.1145s1518701.1518762,10.1145s2998181.2998209,10.1145s2470654.2470714,10.1145s2858036.2858146} & \citepreview{10.1145s3322276.3322296,10.1145s3025453.3025943} \\

Organisation and workplace &
``[- -] will look at historical patterns of work organization and management strategies. It will contrast user-centered concepts of cooperative work, with the idea of seeing cooperative work in the context of \underline{democracy} in the workplace. [- -] The article uses the Scandinavian tradition, with its roots in a Labor Process Approach as a way to analyze the meaning of cooperation for workplace \underline{democracy} and its implication for the design of computer support'' \citep{10.1145s62266.62275},
``[- -] ``As well, we found that computerization of the nursing data led to a shift in the \underline{politics} of the information itself -- the nurses no longer had a cohesive agreement about the kinds of data to enter into the system.'' \citep{10.1145s2702123.2702283}, and
``[- -] we argue that power and \underline{politics} also characterize OSS development [- -]'' \citep{10.1145s192844.192869}
& \citepreview{10.1145s1518701.1519014,10.1145s2818048.2820031,10.1145s2818048.2819952,10.1145s1165387.275650,10.1145s192844.192869,10.1145s800049.801757,10.1145s2702123.2702441,10.1145s2160881.2160886,10.1145s1753326.1753423} & \citepreview{10.1145s3313831.3376284,10.1145s3340631.3394851,10.1145s62266.62275,10.1145s3173574.3173874,10.1145s1013115.1013136,10.1145s1935701.1935773}\\

\multicolumn{4}{l}{\textbf{Methodology}} \\

Design methods &
``[- -] conflicting paradigms are embedded in the legitimization practices of HCI in the \underline{political} realities of computer science and corporate settings leading to contradictions and compromises.'' \citep{10.1145s2858036.2858151} and ``that designers committed to advancing justice and other non-market values must attend not only to the design of objects, processes, and situations, but also to the wider economic and cultural imaginaries of design as a social role. [- -] We argue that designers' elevated status as workers in knowledge economies can have practical consequences for the politics of their design work. [- -]'' \citep{10.1145s2858036.2858592}.
``Often, the multidisciplinary design of business and interactive system concepts is not particularly collaborative nor nearly as user-centered as the organization doing the design claims. [- -] We describe what we did and why and how well what we did worked with particular attention to the affects of organizational culture and \underline{politics} on success.'' \citep{10.1145s778712.778743}, and
``[- -] These activities were created to help a designer collaborate through a more \underline{democratic} process and allow for the creativity of the underrepresented to surface when dealing with complex issues. [- -]'' \citep{10.1145s1640233.1640270}
& \citepreview{10.1145s1378773.1378786,10.1145s2858036.2858482,10.1145s3290605.3300492,10.1145s3078072.3079725,10.1145s3025453.3025535,10.1145s2702123.2702176,10.1145s2468356.2468739,10.1145s3027063.3052760,10.1145s3025453.3025948,10.1145s3290605.3300735,10.1145s1978942.1979003,10.1145s3170427.3174357,10.1145s3025453.3025864,10.1145s778712.778743,10.1145s3313831.3376515,10.1145s1979742.1979671,10.1145s3357236.3395530,10.1145s3173574.3173896,10.1145s3025453.3026031,10.1145s2556288.2557261,10.1145s3357236.3395451,10.1145s2901790.2901861,10.1145s2702123.2702491,10.1145s2858036.2858151,10.1145s2858036.2858592,10.1145s1394445.1394467,10.1145s3290605.3300711,10.1145s1054972.1055076,10.1145s2493432.2493499,10.1145s782896.782909} 
& \citepreview{10.1145s2702123.2702176,10.1145s3294109.3295641,10.1145s1640233.1640270,10.1145s3313831.3376805,10.1145s2556288.2557359} \\

Data analysis methods &
`information retrieval techniques to represent tweets and users as collections of news topics, including high-level categories (e.g., sports, \underline{politics}, business) and detailed subtopics (e.g., Chicago Bulls, Mitt Romney, entrepreneurship).'' \citep{10.1145s2365934.2365943},
``fake news spans almost every realm of human activity, across diverse fields such as \underline{politics} and healthcare'' and developing ``[- -] a graph-based approach for [fake news detection] which operates in three phases'' \citep{10.1145s3372923.3404783}, and 
``Most users on social media have intrinsic characteristics, such as interests and \underline{political} views, that can be exploited to identify and track them.'' \citep{10.1145s3148330.3148340}
& \citepreview{10.1145s2645710.2645740,10.1145s800275.810926,10.1145s2531602.2531636,10.1145s2818346.2820732,10.1145s2645710.2645734,10.1145s2998181.2998259,10.1145s1357054.1357101,10.1145s3148330.3148340,10.1145s800275.810930,10.1145s800275.810928,10.1145s3209542.3209577,10.1145s2365934.2365943,10.1145s3209542.3209549,10.1145s3301275.3302324,10.1145s3372923.3404783,10.1145s2998181.2998254} & \citep{10.1145s3025453.3025885} \\

\multicolumn{4}{l}{\textbf{Technology and society}} \\
Professional topics &
`for decades, HCI scholars have studied technological systems and their relationship to particular contexts and user groups. [- -] Drawing on critical theories, we analyze how the \underline{politics} of titling at CHI functions to build categories of normal and exotic. We explicate the problems that the current ways of representation bring to knowledge production at CHI and necessary paths to move forward.'' \citep{10.1145s3170427.3188409},
''[- -] an explicit \underline{political} vision of an HCI grounded in emancipatory autonomy-an anarchist HCI, aimed at dismantling all oppressive systems by mandating suspicion of and a reckoning with imbalanced distributions of power. We outline some of the principles and accountability mechanisms that constitute an anarchist HCI.'' \citep{10.1145s3290605.3300569}, and
``HCI as a field comfortably and unquestionably links itself with the corporate world. [- -] By undertaking a close reading of a recent publication of a major corporate research lab, I examine what important social and \underline{political} aspects are missing from their vision of the future.'' \citep{10.1145s2851581.2856470}.
& \citepreview{10.1145s3170427.3188396,10.1145s3173574.3174026,10.1145s1520340.1520361,10.1145s3290605.3300569,10.1145s3173574.3173864,10.1145s2559206.2578880,10.1145s3334480.3381808,10.1145s2858036.2858181,10.1145s2556288.2557311,10.1145s1958824.1958827,10.1145s3027063.3052764,10.1145s192844.193064,10.1145s3170427.3188409,10.1145s1457199.1457216,10.1145s2466627.2485921,10.1145s3290605.3300327,10.1145s276758.276778,10.1145s3170427.3188400,10.1145s276755.276762} & \citep{10.1145s1457199.1457216} \\
Technology democratisation &
``a communally accessible, stand-alone voice interface which \underline{democratizes} family access to technology.'' \citep{10.1145s3313831.3376344} and ''[for] visualization [--] to be democratized, we need to provide means for non-experts to create visualizations that allow them to engage directly with datasets. We present constructive visualization a new paradigm for the simple creation of flexible, dynamic visualizations'' \citep{10.1145s2598510.2598566} 
``[- -] We conclude with reflections on BI in the age of AI, big data, and \underline{democratized} access to data analytics.'' \citep{10.1145s1753846.1753872}
& \citepreview{10.1145s2493190.2493205,10.1145s3377325.3377538,10.1145s3170427.3174367,10.1145s2076354.2076393,10.1145s2598510.2598566,10.1145s2540930.2540933,10.1145s2470654.2481360,10.1145s2370216.2370293,10.1145s2470654.2481350,10.1145s3334480.3375220,10.1145s3334480.3383180,10.1145s3332165.3347909,10.1145s1709886.1709985,10.1145s3359996.3364729,10.1145s3313831.3376344,10.1145s3301275.3308447,10.1145s2858036.2858059,10.1145s2675133.2675195,10.1145s2069618.2069665,10.1145s3173574.3173834,10.1145s3345645.3351103} \\

\multicolumn{4}{l}{\textbf{Other}} \\
Motivating context or background &
``[- -] Since non-anthropomorphic avatars are persistent characters engaged in a prolonged performance in virtual worlds, their use also may motivate emerging social mores, \underline{politics} and ideologies. [- -]'' \citep{10.1145s1501750.1501837} and ``[- -] This experiment aims to illustrate the value of a more socially- and location-relevant integration of public displays in our urban neighborhoods as a multifaceted yet \underline{democratic} medium of public communication.'' \citep{10.1145s2491568.2491595}  & \citepreview{10.1145s1394445.1394480,10.1145s985921.986009,10.1145s1501750.1501837,10.1145s1978942.1979290,10.1145s3313831.3376763,10.1145s2501988.2501993,10.1145s2145204.2145220,10.1145s3290605.3300474,10.1145s3371300.3383337,10.1145s3024969.3024971,10.1145s2531602.2531692,10.1145s317456.317469,10.1145s2157689.2157798,10.1145s3313831.3376850,10.1145s1620545.1620571,10.1145s2470654.2470659,10.1145s357744.357919,10.1145s2556288.2557196,10.1145s2531602.2531607,10.1145s1111449.1111476,10.1145s1753326.1753406,10.1145s1460563.1460609,10.1145s1753326.1753610,10.1145s3385956.3418955,10.1145s3313831.3376308,10.1145s2909132.2909249,10.1145s2702123.2702324,10.1145s2207676.2208378,10.1145s3313831.3376193,10.1145s3025453.3025647,10.1145s3125739.3125758,10.1145s2675133.2675195,10.1145s1460563.1460581,10.1145s2935334.2935352,10.1145s2818048.2819942,10.1145s1841853.1841876,10.1145s2858036.2858472,10.1145s3313831.3376167,10.1145s2702123.2702522,10.1145s1056224.1056241,10.1145s3025453.3025983,10.1145s3136755.3136775,10.1145s2145204.2145244,10.1145s99332.99367,10.1145s2998181.2998235,10.1145s3290605.3300586} & \citepreview{10.1145s276755.276783,10.1145s2370216.2370221,10.1145s2491568.2491595,10.1145s3025171.3025227,10.1145s365024.365301,10.1145s3317697.3323358,10.1145s3078072.3079749,10.1145s3173574.3173860} \\
Academic community activities & 
``Since the Arab Spring, there has been a global growing interest in the use of technology in the Arab world. [- -] This one-day workshop aims to bring together HCI researchers/practitioners from the Arab World with those who are conducting/interested in research in this context.'' \citep{10.1145s3064857.3064860} and ``CSCW, like many other academic communities, is reckoning with its roles, responsibilities, and practices amidst 2020's multiple pandemics of COVID-19, anti-Black racism, and a global economic crisis. [- -] This workshop will elicit narratives of good and bad design and data work with communities, apply the lenses of equitable participatory design and data feminism to current CSCW projects and our global context, and develop practical outputs for supporting academics and practitioners in pursuit of \underline{democratic} and just partnerships.'' \citep{10.1145s3406865.3418588}.  & \citepreview{10.1145s3024969.3025041,10.1145s2800835.2812801,10.1145s3393914.3395846,10.1145s2663204.2669986,10.1145s3148330.3152697,10.1145s3022198.3026317,10.1145s3345645.3351105,10.1145s3270316.3270594,10.1145s2851581.2886426,10.1145s2675133.2697079,10.1145s2559206.2559974,10.1145s3301019.3319993,10.1145s3064857.3079113,10.1145s3267305.3274189,10.1145s2818052.2869132,10.1145s2858036.2858420,10.1145s1753846.1753898,10.1145s2957276.2996281,10.1145s3290607.3299033,10.1145s3406865.3418593,10.1145s1979742.1979869,10.1145s3301019.3325148,10.1145s3379350.3416154,10.1145s3301019.3319994,10.1145s3022198.3024945,10.1145s3170427.3173027,10.1145s3301019.3324872,10.1145s2638728.2641309,10.1145s3170427.3173024,10.1145s2818052.2855505,10.1145s3170427.3188550,10.1145s3406865.3418562,10.1145s3311957.3359438,10.1145s3294109.3305257,10.1145s2702613.2727687,10.1145s2908805.2913020,10.1145s2702613.2702657,10.1145s3311957.3361853,10.1145s3064857.3079111,10.1145s3064857.3064860,10.1145s1520340.1520552,10.1145s2677199.2691608,10.1145s2702613.2702616,10.1145s997078.997117,10.1145s1499224.1499226,10.1145s2851581.2886436,10.1145s2441776.2441779,10.1145s2685553.2699325,10.1145s3294109.3301262,10.1145s2598784.2598795,10.1145s2793107.2810283,10.1145s2745197.2755521,10.1145s2957276.2996292,10.1145s3027063.3053265,10.1145s2818052.2874318,10.1145s3334480.3375180,10.1145s3272973.3274079,10.1145s3334480.3381055,10.1145s1979742.1979543,10.1145s3372923.3404477,10.1145s2494091.2499623,10.1145s2909132.2927472,10.1145s3170427.3186486,10.1145s2559206.2559971,10.1145s1753846.1753940,10.1145s1979742.1979568,10.1145s2598784.2611381,10.1145s3123021.3123073,10.1145s2494091.2496042,10.1145s3345645.3351106,10.1145s1520340.1520731,10.1145s3027063.3049280,10.1145s2212776.2212715,10.1145s1520340.1520574,10.1145s3170427.3185363,10.1145s3272973.3274064,10.1145s2793107.2810259,10.1145s3027063.3027138,10.1145s2800835.2804403,10.1145s3393914.3395858,10.1145s2851581.2891097,10.1145s3209542.3209543,10.1145s3059454.3059495,10.1145s3272973.3272977,10.1145s1358628.1358925,10.1145s2818052.2874354,10.1145s3393914.3395864,10.1145s3027063.3056455,10.1145s1125021.1125072,10.1145s2851581.2856470,10.1145s1753846.1753870,10.1145s2212776.2223843,10.1145s1031607.1031618,10.1145s1979742.1979835,10.1145s3383668.3419927,10.1145s1841853.1841906,10.1145s765891.765934,10.1145s2800835.2804399,10.1145s3272973.3272983,10.1145s2141512.2141601,10.1145s3290607.3313061,10.1145s3059454.3059485,10.1145s3311957.3361852,10.1145s3272973.3274040,10.1145s1753846.1754086,10.1145s3064857.3079141,10.1145s3342220.3344782,10.1145s1125451.1125762,10.1145s2851581.2886427,10.1145s2638728.2641685,10.1145s2851581.2856499,10.1145s3258859,10.1145s3255629,10.1145s3251757,10.1145s3242671.3242715,10.1145s3334480.3386149,10.1145s2468356.2479538,10.1145s3406865.3418585,10.1145s2702613.2702617,10.1145s1520340.1520446,10.1145s3397617.3397834,10.1145s2559206.2559227,10.1145s2559206.2581358,10.1145s3295750.3298944,10.1145s3131785.3131845,10.1145s3342220.3344784,10.1145s3372923.3404476,10.1145s3406865.3418372,10.1145s2818052.2869111,10.1145s3311957.3359484,10.1145s2559206.2574795,10.1145s354384.354390,10.1145s3311957.3359493,10.1145s1753846.1753906,10.1145s3290607.3299008,10.1145s3325480.3326550,10.1145s2145204.2145206,10.1145s1165387.275652,10.1145s3301019.3323887,10.1145s1622176.1622211,10.1145s3345645.3351103,10.1145s1864431.1864502,10.1145s3290607.3299010,10.1145s2389176.2389233,10.1145s3311957.3359488,10.1145s1056808.1056940,10.1145s3406865.3418311,10.1145s3272973.3274062,10.1145s506443.506649,10.1145s2494091.2499225,10.1145s2556420.2556511,10.1145s3294109.3302935,10.1145s3027063.3050431,10.1145s3322276.3325420,10.1145s2818052.2893359,10.1145s3170427.3180647,10.1145s1056808.1057014,10.1145s765891.765937,10.1145s634067.634129,10.1145s3334480.3375157} & \citepreview{10.1145s2818052.2869125,10.1145s3383668.3419873,10.1145s3270316.3270320,10.1145s3022198.3024948,10.1145s3379336.3381474,10.1145s3342220.3344930,10.1145s3173574.3173583,10.1145s2800835.2809508,10.1145s3406865.3418562,10.1145s2968219.2971435,10.1145s3170427.3186539,10.1145s3379337.3422877,10.1145s3170427.3186478,10.1145s2702613.2702657,10.1145s3064857.3064860,10.1145s2441776.2441779,10.1145s2660398.2660435,10.1145s2607023.2611454,10.1145s1753846.1753936,10.1145s2793107.2810283,10.1145s2745197.2755521,10.1145s2876456.2879482,10.1145s3248982,10.1145s3027063.3056455,10.1145s1520340.1520472,10.1145s3334480.3375179,10.1145s3294109.3305257,10.1145s1979742.1979493,10.1145s2757226.2757369,10.1145s2702613.2702656,10.1145s3027063.3049280,10.1145s997078.997092,10.1145s1640233.1640309,10.1145s3290607.3313771,10.1145s1640233.1640302,10.1145s1520340.1520574,10.1145s3170427.3185363,10.1145s2212776.2212427,10.1145s2451176.2451230,10.1145s1958824.1958929,10.1145s3311957.3359486,10.1145s3313831.3376409,10.1145s1358628.1358925,10.1145s2559206.2579405,10.1145s1935701.1935758,10.1145s1979742.1979852,10.1145s2069618.2069637,10.1145s1031607.1031618,10.1145s2851581.2912563,10.1145s3383668.3419927,10.1145s2851581.2912558,10.1145s3170427.3188671,10.1145s2141512.2141601,10.1145s3311957.3361852,10.1145s2556420.2556785,10.1145s3349263.3351332,10.1145s2800835.2801662,10.1145s3334480.3381068,10.1145s3406865.3418372,10.1145s3290607.3312892,10.1145s3311957.3359484,10.1145s1125451.1125517,10.1145s1622176.1622211,10.1145s2957276.2997020,10.1145s3406865.3418588,10.1145s3334480.3375035,10.1145s3311957.3359488,10.1145s1056808.1056940,10.1145s2638728.2641681,10.1145s2494091.2499225,10.1145s985921.986190,10.1145s2468356.2468837} \\

\end{longtable}

\newpage

\section{Literature reviewed}
\label{literature_reviewed}

\bibliographystylereview{ACM-Reference-Format}
\bibliographyreview{reviewall,reviewall2}

\end{document}